\documentclass[fleqn,usenatbib]{mnras}

\usepackage[T1]{fontenc}
\usepackage{ae,aecompl}

\usepackage{graphicx}	% Including figure files
\usepackage{amsmath}	% Advanced maths commands
\usepackage{amssymb}	% Extra maths symbols
\usepackage{times}

\makeatother
\usepackage{newtxtext,newtxmath}
\title[Synthetic observations of the Brick]{The complex multi-scale structure in simulated and observed emission maps of the proto-cluster cloud G0.253+0.016 (`the Brick')}

\author[M. A. Petkova et al.]{Maya~A.~Petkova,$^{1,2}$\thanks{E-mail: maya.petkova@chalmers.se (MAP)}
J.~M.~Diederik~Kruijssen,$^{1,3}$
A.~Louise~Kluge,$^{1}$
\newauthor
Simon~C.~O.~Glover,$^{4}$
Daniel~L.~Walker,$^{5}$
Steven~N.~Longmore,$^{6}$
\newauthor
Jonathan~D.~Henshaw,$^{6,7}$
Stefan Reissl,$^{4}$
and James~E.~Dale$^{8}$
\\
$^{1}$Astronomisches Rechen-Institut, Zentrum f{\"u}r Astronomie der Universit{\"a}t Heidelberg, M{\"o}nchhofstra{\ss}e 12-14, D-69120 Heidelberg, Germany\\
$^{2}$Space, Earth and Environment Department, Chalmers University of Technology, SE-412 96 Gothenburg, Sweden\\
$^{3}$Cosmic Origins Of Life (COOL) Research DAO, coolresearch.io\\
$^{4}$Universit{\"a}t Heidelberg, Zentrum f{\"u}r Astronomie, Institut f{\"u}r Theoretische Astrophysik, Albert-Ueberle-Str 2, D-69120 Heidelberg, Germany\\
$^{5}$Department of Physics, University of Connecticut, 196A Auditorium Road, Storrs, CT 06269 USA\\
$^{6}$Astrophysics Research Institute, Liverpool John Moores University, IC2, Liverpool Science Park, 146 Brownlow Hill, Liverpool L3 5RF, UK\\
$^{7}$Max-Planck-Institut f{\"u}r Astronomie, K{\"o}nigstuhl 17, D-69117, Heidelberg, Germany\\
$^{8}$Centre for Astrophysics Research, University of Hertfordshire, Hatfield, AL10 9AB, UK
}

\date{Accepted XXX. Received 2022 September 3; in original form 2021 April 19}

\pubyear{2021}

\begin{document}
\label{firstpage}
\pagerange{\pageref{firstpage}--\pageref{lastpage}}
\maketitle

\begin{abstract}
The Central Molecular Zone (CMZ; the central $\sim500$~pc of the Milky Way) hosts molecular clouds in an extreme environment of strong shear, high gas pressure and density, and complex chemistry. G0.253+0.016, also known as `the Brick', is the densest, most compact and quiescent of these clouds. High-resolution observations with the Atacama Large Millimeter/submillimeter Array (ALMA) have revealed its complex, hierarchical structure. In this paper we compare the properties of recent hydrodynamical simulations of the Brick to those of the ALMA observations. To facilitate the comparison, we post-process the simulations and create synthetic ALMA maps of molecular line emission from eight molecules. We correlate the line emission maps to each other and to the mass column density, and find that HNCO is the best mass tracer of the eight emission lines within the simulations. Additionally, we characterise the spatial structure of the observed and simulated cloud using the density probability distribution function (PDF), spatial power spectrum, fractal dimension, and moments of inertia. While we find good agreement between the observed and simulated data in terms of power spectra and fractal dimensions, there are key differences in the density PDFs and moments of inertia, which we attribute to the omission of magnetic fields in the simulations. 
This demonstrates that the presence of the Galactic potential can reproduce many cloud properties, but additional physical processes are needed to fully explain the gas structure.
\end{abstract}


%%%%%%%%%%%%%%%%%%%%%%%%%%%%%%%%%%%%%%%%%%%%%%%%%%

%%%%%%%%%%%%%%%%% BODY OF PAPER %%%%%%%%%%%%%%%%%%

\section{Introduction}
G0.253+0.016, also known as `the Brick', is a massive ($\sim 10^5~\mathrm{M}_{\odot}$), compact (2--3~pc) cloud in the Central Molecular Zone (CMZ) of the Milky way \citep{Lis1994,Longmore2012,Kauffmann2013,Rathborne2015,Nogueras-Lara2021}. The Brick belongs to a stream of clouds on an eccentric orbit around the Galactic Centre, with an orbital radius varying between 60 and 100~pc \citep{Bally1987,Molinari2011, Kruijssen2015,Henshaw2016}. Together with the rest of the CMZ clouds, the Brick is subjected to strong shear \citep[$V/R \sim 1.7~\mathrm{Myr}^{-1}$;][]{Kruijssen2014,Krumholz2015,Krumholz2017,Federrath2016,Jeffreson2018}, which has been argued to drive solenoidal turbulence \citep{Federrath2016,Kruijssen2019}. Additionally, the Brick experiences an unusually compressive tidal field, acting to promote collapse \citep{Kruijssen2015}. These two external influences likely play a key role in the cloud's observed high velocity dispersion and high column density \citep{Kruijssen2019}.

However, despite its high density, the Brick exhibits a very low star formation rate (SFR), which is a long-standing puzzle \citep[e.g.][]{Longmore2012,Rathborne2014b,Henshaw2019}. The same applies to the CMZ as a whole \citep{Longmore2013,Barnes2017}, even though some individual clouds, such as Sgr B2, are actively forming stars \citep{Ginsburg2018} and stellar clusters \citep{Ginsburg2018b}. One explanation for the presently low SFR in the region is the idea that the CMZ undergoes bursts of  star formation, followed by quiescent phases \citep{Kruijssen2014,Krumholz2017,Armillotta2020,Sormani2020,Tress2020}. \citet{Longmore2013b} and \citet{Barnes2019,Barnes2020} reported an increase of star formation activity as a function of orbital position angle among the subset of CMZ clouds known as the `dust ridge' (which includes the Brick). This idea implies that the `dust ridge' clouds may follow an evolutionary sequence, where the Brick is positioned on the cusp of starting to form stars and potentially a young massive cluster. Indeed, recent work by \citet{Walker2021} and \citet{Henshaw2022} found signatures of early star formation within the cloud.
Therefore, the Brick represents a unique opportunity for studying the initial developmental stages of a likely progenitor of a massive stellar cluster, similar to the Arches \citep{Rathborne2015}. Furthermore, the Brick can help us understand star formation in extreme environments, such as present-day galactic centres and high-redshift galaxies (which share similar pressures and temperatures), without the resolution restrictions present in 
studies of objects at so much greater distances \citep{Kruijssen2013}.

The structure of the Brick has been studied with the Atacama Large Millimeter/submillimeter Array (ALMA) \citep{Rathborne2015}. These high-resolution  observations revealed complex, hierarchical substructure, in what was previously seen as a smooth, dark cloud \citep[also see][]{Henshaw2019}. \citet{Rathborne2015} detected the line emission from 17 different molecules, which trace different chemistry and physical processes within the Brick. They report high fractal dimensions and large velocity dispersions for all of their observations, which is consistent with a highly supersonically turbulent environment. 

Additionally, there have been numerical efforts towards modelling the gas dynamics and star formation in the CMZ. Some authors have performed simulations that follow the gas flow towards the galactic centre and connect kiloparsec scales to the inner few hundred parsecs \citep[e.g.][]{Sormani2015,Armillotta2019,Armillotta2020,Tress2020,Sormani2020, Moon2021}. These simulations allow us to study the formation and large-scale properties of molecular clouds in the CMZ. However, they do not yet have enough resolution to model the small-scale structures within the individual clouds. For this reason, it is important to also consider simulations of individual clouds within the CMZ environment, as done recently by \citet{Dale2019} and \citet{Kruijssen2019}. These works have focused on individual CMZ clouds moving in an external gravitational potential chosen to closely represent the potential in the CMZ.

To determine the effect of the Galactic potential on the star formation process in the clouds, \citet{Dale2019} performed a set of isolated cloud simulations and compared them to several sets of simulations evolved on orbits through the external potential. They found that the isolated simulations had very different SFR, sizes and velocity dispersions in comparison to the simulations evolved in the potential. This indicates that the Galactic potential has an important role in shaping the CMZ clouds.

In addition, \citet{Dale2019} used two sets of initial conditions for the simulations evolved with an external gravitational potential. Previous cloud-scale simulations \citep[e.g.][]{Bertram2015,Federrath2016} created initial conditions by selecting a virial parameter (ratio of kinetic to gravitational potential energy) of the cloud. However, the gravitational potential adopted by \citet{Dale2019} generates a compressive tidal field at the clouds' orbits \citep{Kruijssen2015}, implying that their overall behaviour and evolution cannot be predicted using the virial parameter of a cloud in isolation. \citet{Dale2019} constructed initial conditions in which the clouds have turbulent support against the compressive tidal field, and not only against their own self-gravity. In their work, they demonstrated that this `tidally-virialised' set of clouds is a better match for the observed SFR in the CMZ than classical `self-virialised' clouds, as the latter underwent rapid gravitational collapse, inconsistent with the observations of the Brick. Additionally, \citet{Kruijssen2019} found that the tidally-virialised simulations reproduce known CMZ properties, such as the cloud inclinations in the plane of the sky, aspect ratios, velocity dispersion, column densities and retrograde line of sight velocity gradients.

%The aim of this work is to provide further comparison between a tidally-virialised and a self-virialised simulation of the Brick \citep{Dale2019,Kruijssen2019} by studying the substructure of the clouds and comparing the simulated substructure with the ALMA observations of \citet{Rathborne2015}. 
The aim of this work is to investigate to what extent the influence of the Galactic potential may be responsible for shaping the CMZ gas on sub-cloud scales. We achieve this by comparing the cloud-scale simulations of the Brick \citep{Dale2019,Kruijssen2019} and the ALMA observations of \citet{Rathborne2015}. For completeness, we use both the tidally-virialised and  self-virialised simulations, and characterise their substructure.
In order to 
%achieve this goal
facilitate the comparison, we post-process a snapshot of each simulation and create synthetic ALMA emission maps. The exact steps for generating the synthetic observations are discussed in Section~\ref{sec:methods}. We then study the structure of the emission maps in Section~\ref{sec:comparison}, and perform a quantitative comparison to the observations in Section~\ref{sec:cloud-structure}. Finally, our main conclusions are summarised in Section~\ref{sec:conclusion}.

\section{Methods}
\label{sec:methods}
\subsection{SPH simulations}
\label{sec:SPH} 

In this work we have used two of the hydrodynamics simulations presented in \citet{Dale2019} and have created synthetic emission maps based on these. We have selected one snapshot per simulation, corresponding to the present day position of the real Brick cloud. The simulations were performed with the SPH (smoothed particle hydrodynamics) code \textsc{gandalf} \citep{Hubber2018}. The chosen simulations are the high-density, tidally-virialised model (TVir) and the high-density, self-virialised model (SVir) from \citet{Dale2019}. The former was used as an analogue to the Brick in the analysis of \citet{Kruijssen2019}, due to its approximately appropriate size, surface density, and velocity dispersion. The simulated clouds both have the same initial gas mass of approximately $4.5\times 10^5$ M$_{\sun}$, equally divided between $10^6$ SPH particles. Both clouds were selected from a set of randomly generated velocity fields to have negative spin angular momentum consistent with the shearing motions in the clouds observed upstream from the Brick, and have been evolved along the eccentric orbit of \citet{Kruijssen2015} and \citet{Henshaw2016}. Along their orbit, the clouds have been experiencing their own self-gravity, as well as the static external gravitational potential, derived from the observed stellar distribution in the CMZ \citep{Launhardt2002}. The difference between the two simulations is in their initial velocity dispersion. While the SVir model has been assigned velocities based on an assumed internal virial parameter, the TVir model's velocities have been chosen to balance out the compressive tidal field, caused by the external potential. As a result, the SVir model is under-supported against the tidal forces and it undergoes more rapid collapse \citep{Dale2019}. While the clouds have been allowed to collapse and form sink particles, corresponding to sites of star formation, stellar feedback has been omitted from the simulations in order to isolate the effects of the external potential and the eccentric orbit. 

The density threshold used for the sink creation is $\rho_{\rm sink}=10^{-17} \mathrm{g~cm}^{-3}$, which leads to sink accretion radii $R_{\rm sink} = 0.035 \mathrm{pc}$. The chosen threshold density is lower than the maximum resolvable density ($1.9 \times 10^{-17} \mathrm{g~cm}^{-3}$), as given by the \citet{Bate1997} criterion, which ensures numerical accuracy of the behaviour of gas near the Jeans limit. By the time the simulated clouds reach the present-day position of the Brick along its orbit (after 0.74~Myr of evolution), $\sim 55\%$ (TVir) and $\sim 65\%$ (SVir) of their gas mass is transformed into sink particles. Note that at the particle resolution of these simulations ($m_{\rm part}=0.4463$~M$_{\odot}$) the sinks do not represent individual stars, but only dense clumps that have collapsed\footnote{Each sink particle contains tens of gas particles. Therefore, with the particle mass being $m_{\rm part}=0.4463$~M$_{\odot}$, the lightest sink has a mass much greater than 1~M$_{\odot}$}. For the subsequent analysis in this paper we ignore the sinks and only consider the remaining gas particles. The gas has been evolved using an isothermal equation of state, with a temperature of 65~K, which is a typical value observed in the CMZ \citep{Ao2013,Ginsburg2016,Krieger2017}. The assumed mean molecular weight is $\mu=2.35$.

In their initial state, the clouds have been assigned a velocity field, consistent with turbulent motion, with a power spectrum of the form $P(k) \propto k^{-4}$, and resulting in virial parameters $\alpha_{vir} = 3.2$ (TVir) and $\alpha_{vir} = 2.6$ (SVir). These values later drop to 1.24 and 1.16 for TVir and SVir, respectively, by the time the clouds reach the present-day position of the Brick (which corresponds to the analysed snapshots in this paper). Even though there is no explicit subsequent driving of turbulence, the presence of the strong shear in the external gravitational potential maintains a large velocity dispersion \citep{Kruijssen2019}.

\subsection{Post-processing work flow}

\begin{figure}
	\includegraphics[width=\columnwidth]{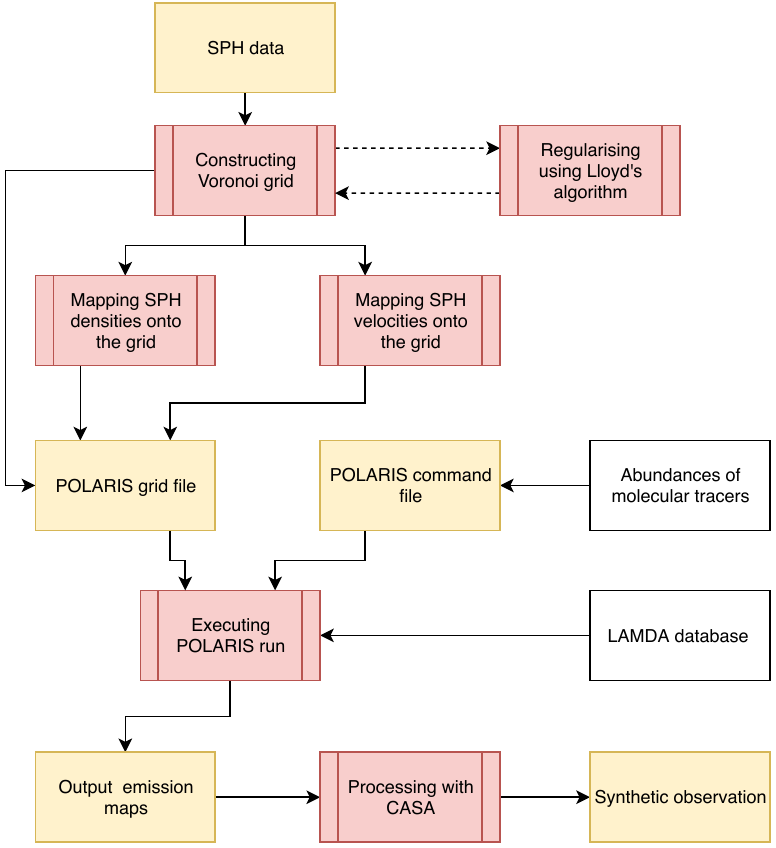}
    \caption{A schematic of the post-processing work flow. The SPH data are used for the construction of a Voronoi grid, which is regularised with Lloyd's algorithm \citep{Lloyd1982}. The SPH densities and velocities are then mapped onto the grid and, together with the grid's neighbours lists, are combined into a \textsc{polaris} grid file. We additionally create a \textsc{polaris} command file, which contains our assumed abundances of molecular tracers (see Table~\ref{tab:abundances}). \textsc{polaris} is executed using the grid file, the command file, and information on molecular transitions from the LAMDA database \citep{Schoiler2005}. Finally, the output emission maps from \textsc{polaris} are further processed with the \textsc{casa} package \citep{McMullin2007} to create synthetic ALMA observations.}
    \label{fig:post_process_flow}
\end{figure}

We have created two sets of synthetic ALMA emission maps, one per snapshot of each simulation. The snapshots correspond to the present day orbital position of the Brick. The emission maps are produced as a result of post-processing the snapshots with the radiative transfer code \textsc{polaris} \citep{Reissl2016}, and then further with the \textsc{casa} (Common Astronomy Software Applications) package \citep{McMullin2007}. To facilitate these processing steps, we have made several intermediate steps, as shown in Figure~\ref{fig:post_process_flow}. First, the SPH particle positions of each snapshot have been used in order to generate a Voronoi tesselation grid. The grid has been regularised using Lloyd's algorithm \citep{Lloyd1982} to ensure that the gas structure is well resolved. After the grid construction, SPH properties such as density and velocity have been mapped onto the grid cells, using the mapping method described in \citet{Petkova2018}. All of the above data have been combined to create an input file for \textsc{polaris}, which is used for performing the desired radiative transfer. More detailed descriptions of each of the individual steps of the process are given below.

\subsection{Choice of Voronoi grid}
\begin{figure*}
	\includegraphics[width=\textwidth]{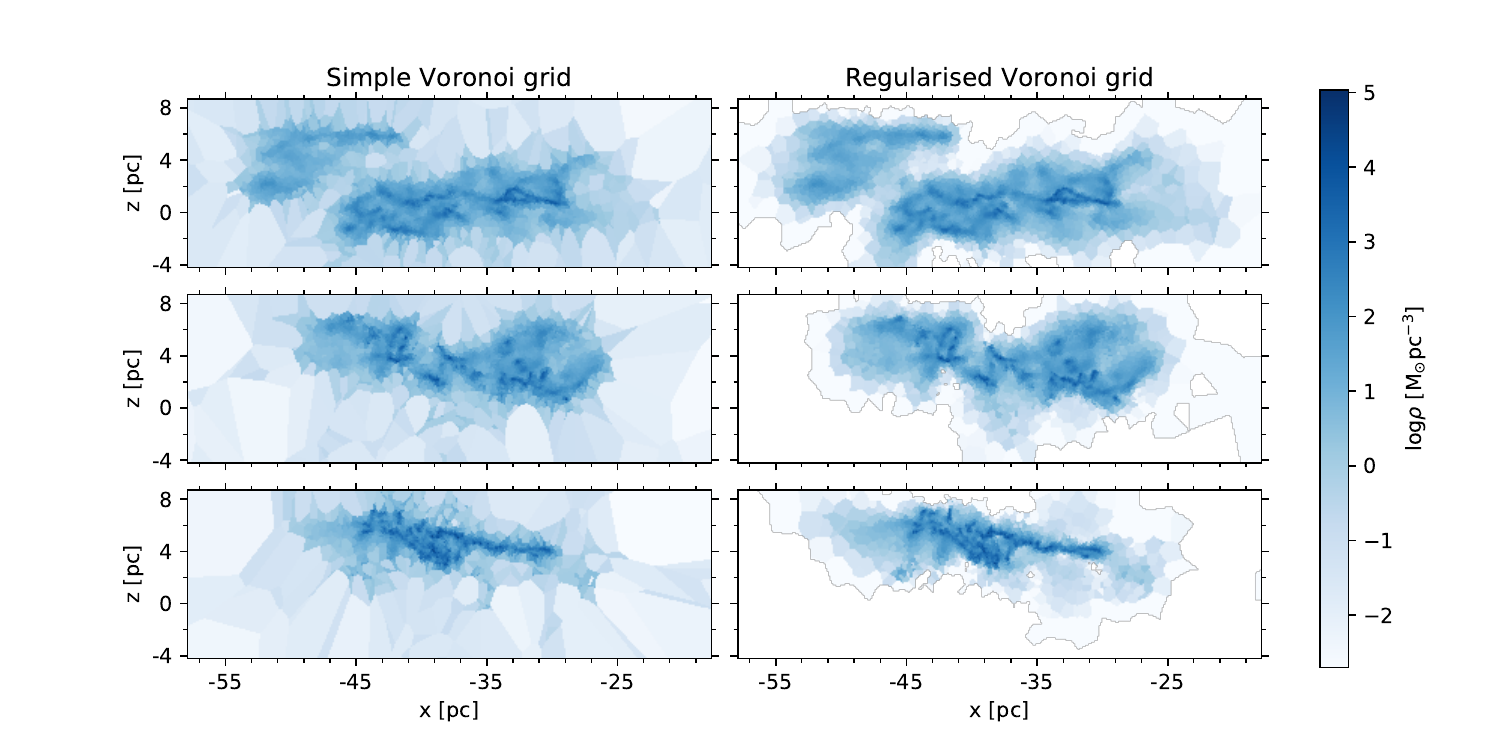}
    \caption{Volume density plots of different slices through the TVir cloud. The slices are in the planes of $y=55$~pc (\textit{top}), $y=62$~pc (\textit{middle}) and $y=65$~pc (\textit{bottom}), with darker blue being higher density. The plots on the left use the SPH particle positions as generating sites, while the plots on the right have grid cells regularised with Lloyd's algorithm with 5 iterations, and represent the cloud structure more cleanly.}
    \label{fig:grids}
\end{figure*}

\begin{figure*}
	\includegraphics[width=\textwidth]{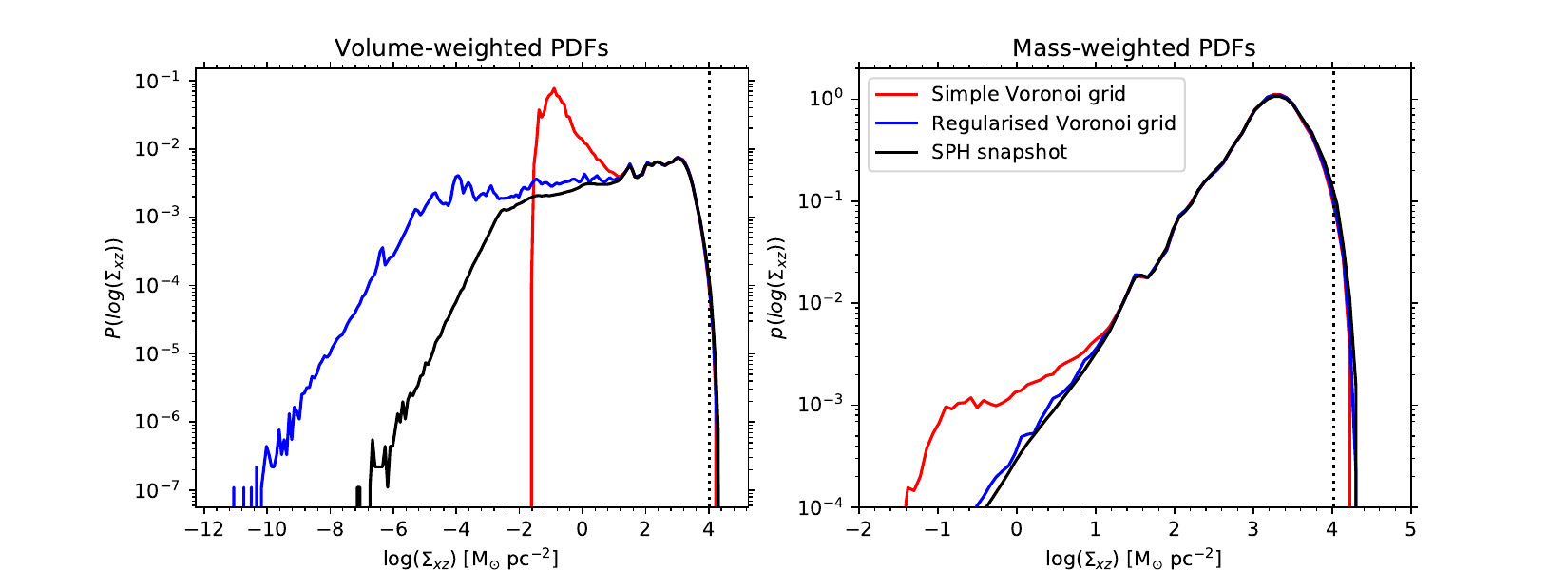}
    \caption{Column density PDFs of the simple Voronoi grid, the Voronoi grid regularised with Lloyd's iterations, and the SPH snapshot of the TVir cloud. The left-hand panel presents the volume-weighted PDFs, and the right-hand panel contains the mass-weighted ones. The three underlying column density maps have been projected onto the plane of the sky, with a resolution of $3008 \times 3008$ pixels, and a pixel size of $0.33^{''}$ (0.013~pc). The vertical dotted lines indicate the approximate equivalent of the volume density threshold for sink formation in the simulations (see the text).}
    \label{fig:dens-pdfs}
\end{figure*}

In order to post-process the simulation snapshots with \textsc{polaris}, it is necessary to use a grid representation of the gas distribution, as the radiative transfer code cannot be executed directly on the SPH particles.

A Voronoi grid is an unstructured mesh built around a set of generating sites. The grid is constructed such that each cell encloses all points in space closer to its own generating site than to any other generating site. This results in all grid cells in 3D having the shape of convex polyhedra \citep{Dirichlet1850,Voronoi1908}.

The benefit of adopting a Voronoi grid to represent SPH data is that a large span of length scales can be covered by a small number of grid cells. A Voronoi grid can follow the SPH particle distribution and hence provide better resolution in regions that require it. The most intuitive way of constructing the grid is, therefore, by using the particle positions as generating sites (which we will refer to as a simple Voronoi grid).

This approach does have some disadvantages, as discussed by \citet{Koepferl2017}. Due to the geometric property of a Voronoi grid to bisect the distance between two neighbouring generating sites, grid cells on the interface between a high and a low density region end up being artificially elongated and make the dataset noisy. The left-hand panels of Figure~\ref{fig:grids} illustrate this issue. We show three density slices through the Voronoi grid of the TVir snapshot (at $y=55,62,65$~pc), which all exhibit sharp edges and low-density "noise" around their periphery. To counter that problem \citet{Koepferl2017} proposed that additional generating sites are inserted at chosen locations. Here we propose an alternative approach, consisting of two steps. The first step is to insert a small number (TVir: 770; SVir: 796) of additional generating sites in the empty areas surrounding the cloud. We have arranged these on a coarse hexagonal grid. This is necessary because \textsc{polaris} requires a cubic simulation volume as an input (see Section~\ref{sec:polaris}), and the simulated clouds have one spatial dimension with much smaller range than the other two. The second step is to use Lloyd's algorithm to regularise the shapes of the grid cells \citep{Lloyd1982}. Lloyd's algorithm follows an iterative procedure. As an initial step, we construct a grid which uses the SPH particle positions as generating sites. We then move each generating site to the centroid of its cell and reconstruct the grid. The repositioning of generating sites and grid reconstruction are repeated for several iterations (here we have chosen to use five, ensuring visual convergence of the density distribution). This process regularises the cell shapes and it produces a clean final look, as can be seen in the right-hand panels of Figure~\ref{fig:grids}. Note that with each iteration the grid loses some information about the SPH particle distribution. Therefore, in the limit of infinite iterations, the grid cells would approach a uniform distribution.

In addition to the visual comparison of the two Voronoi grids of the TVir snapshot in Figure~\ref{fig:grids}, we show the volume-weighted and mass-weighted probability distribution functions (PDFs) of the column density maps of the two grid representations and the original SPH snapshot in Figure~\ref{fig:dens-pdfs}. We only show the PDFs of the TVir snapshot, since those of the SVir snapshot are almost identical. The first thing to notice about the PDFs shown in Figure~\ref{fig:dens-pdfs} is that there is a good agreement between the three density distribution representations for column densities $\Sigma_{xz} \geq 16~\textrm{M}_{\odot}~\textrm{pc}^{-2}$. For lower column densities, however, there are some discrepancies. When considering the volume-weighted PDFs, we see that the shape of the SPH snapshot curve vaguely resembles a log-normal with a flattened tail towards low densities. The PDF peaks at $\log(\Sigma_{xz}/(\textrm{M}_{\odot}~\textrm{pc}^{-2})) \approx 3$, with a rapid decline towards higher densities. Below the peak, there is an approximately flat, but decreasing region, until $\log(\Sigma_{xz}/(\textrm{M}_{\odot}~\textrm{pc}^{-2})) \approx -3$, below which the density PDF decreases more rapidly. A similar trend is followed by the PDF of the regularised Voronoi grid, but the approximately flat region extends to lower column densities ($\log(\Sigma_{xz}/(\textrm{M}_{\odot}~\textrm{pc}^{-2})) \approx -5$). This excess of low-density pixels in the regularised Voronoi grid with respect to the SPH snapshot comes from the fact that some of the zero column density pixels in the SPH column density map have very low but non-zero densities in the column density map of the regularised Voronoi grid. This is the result of a smoothing effect, where a small fraction of a particle's smoothing kernel volume overlaps with a grid cell at the periphery of the cloud, and hence produces a non-zero mass (and density) in the cell.

By contrast, the simple Voronoi grid does not have any pixels with zero column density by construction (since each grid cell contains a particle). As a result, its density PDF has a second, higher peak at $\log(\Sigma_{xz}/(\textrm{M}_{\odot}~\textrm{pc}^{-2})) \approx -1$, and no pixels with $\log(\Sigma_{xz}/(\textrm{M}_{\odot}~\textrm{pc}^{-2})) < -2$.
When considering the mass-weighted column density PDFs, which highlight the densities where the mass is primarily contained, the three representations of the snapshots are considerably more similar. The density PDFs of the SPH distribution and regularised Voronoi grid are both single-peak curves with close to identical shapes. The simple Voronoi grid matches the position of this peak in its PDF, but it contains an additional feature at lower densities, corresponding to the mass contained in its elongated cells at the cloud periphery.

In addition, due to the finite resolution of the simulations, the high-mass end of the column density PDFs decreases more steeply than it should in reality. Gas above the volume density threshold of sink formation is trapped in sink particles and it is not included in the PDFs. We convert this threshold to column density using $\Sigma_{\rm sink}=2 R_{\rm sink} \rho_{\rm sink} \approx 1000~\textrm{M}_{\odot}~\textrm{pc}^{-2}$, and we include it in Figure~\ref{fig:dens-pdfs} as vertical dotted lines.

Together, Figure~\ref{fig:grids} and Figure~\ref{fig:dens-pdfs} demonstrate that the regularised Voronoi grid is a better match of the SPH snapshot than the simple Voronoi grid. Therefore, for the rest of this work we adopt the regularised Voronoi grid. We perform the grid construction using the C++ library Voro++ \citep{Rycroft2009}.

\subsection{Mapping of SPH parameters onto grid cells}

In order to perform the line radiative transfer, we must assign density and three-dimensional velocity to each grid cell of the Voronoi grid. We do this self-consistently by calculating the average cell density and velocity directly from the SPH dataset. This method is an extension of the mapping presented in \citet{Petkova2018} and it can be derived directly from the SPH formalism.

Let us consider a fluid property $A$, which is defined for each particle $b$. We can express $A$ at an arbitrary position $\mathbf{r}$ as:

\begin{equation}
A(\mathbf{r}) \approx \sum_{b=1}^{N} \frac{A_b m_b}{\rho_b} W(\mathbf{r}-\mathbf{r}_b, h_b),
\label{eq:SPH-formalism}
\end{equation}
where $N$ is the total number of particles, $m_b$ and $\rho_b$ are the mass and density of $b$, $\mathbf{r}_b$ is the position vector of $b$, and $W$ is the kernel function. The kernel function is approximately Gaussian-shaped and it equals zero for $|\mathbf{r}-\mathbf{r}_b|$ greater than a certain distance, given as a multiple of the smoothing length $h_b$. The most commonly used SPH kernel function, and the one adopted by this work, is the cubic spline, defined as \citep{Monaghan1985}:

\begin{equation}
   W(r,h) = \frac{1}{h^3\pi}  \begin{cases}
      1 - 1.5 \left( \frac{r}{h} \right )^2 + 0.75 \left ( \frac{r}{h} \right )^3, & r\leq h; \\
      0.25 \left (2 - \left (\frac{r}{h} \right ) \right )^3, & h< r< 2h; \\
      0, & r\geq 2h. \\
  \end{cases} 
  \label{eq:cubic-spline-3d}
\end{equation}
If we now consider a grid cell, the volume-averaged value of $A$ in cell $i$ is given as:

\begin{equation}
A_i =  \frac{\int_{V_i}  A(\mathbf{r'}) \mathrm{d}V'}{\int_{V_i} \mathrm{d}V'} = \frac{1}{V_i} \int_{V_i}  A(\mathbf{r'}) \mathrm{d}V',
\label{eq:average-cell}
\end{equation}
where $V_i$ is the volume of cell $i$. Combining equations ~\ref{eq:SPH-formalism} and ~\ref{eq:average-cell} we can write that:

\begin{eqnarray}
A_i & =  & \frac{1}{V_i} \int_{V_i}  \sum_{b=1}^{N} \frac{A_b m_b}{\rho_b} W(\mathbf{r}-\mathbf{r}_b, h_b) \mathrm{d}V' \\
 & = & \frac{1}{V_i} \sum_{b=1}^{N} \frac{A_b m_b}{\rho_b} \int_{V_i} W(\mathbf{r}-\mathbf{r}_b, h_b) \mathrm{d}V'.
\end{eqnarray}
Using a volume-averaged value is appropriate for performing the density mapping, because that ensures mass conservation between the SPH snapshot and the Voronoi grid. Hence we have the following expression for the cell density, $\rho_i$:
\begin{equation}
\rho_i =  \frac{1}{V_i} \sum_{b=1}^{N} m_b \int_{V_i} W(\mathbf{r}-\mathbf{r}_b, h_b) \mathrm{d}V'.
\label{eq:cell-dens}
\end{equation}
Note that in the above expression, the sum gives the cell mass, which is comprised of contributions from different particles.

In contrast to the density mapping, the velocity mapping is typically done as a mass-average (since that conserves momentum). Therefore, we will use the following expression for each velocity component, $v_{j,i}$ ($j=\{1,2,3\}$):
\begin{equation}
v_{j,i} =  \frac{\sum_{b=1}^{N} v_{j,b} m_b \int_{V_i} W(\mathbf{r}-\mathbf{r}_b, h_b) \mathrm{d}V'}{\sum_{b=1}^{N} m_b \int_{V_i} W(\mathbf{r}-\mathbf{r}_b, h_b) \mathrm{d}V'}.
\label{eq:cell-vels}
\end{equation}

By using the equations \ref{eq:cell-dens} and \ref{eq:cell-vels}, we reduce the density and velocity mapping to simple sums, in which the only remaining unknown is the integral of the kernel function. Calculating this integral numerically is computationally expensive, hence we use the analytic form of the integral derived in \citet{Petkova2018}, and we include a summary of it in Appendix ~\ref{sec:app-density}.

\subsection{Processing with \textsc{polaris}}
\label{sec:polaris}

We perform radiative transfer post-processing of the simulated Brick with \textsc{polaris} \citep{Reissl2016,Brauer2017,Reissl2019}. The code uses 3D Monte Carlo radiative transfer for dust scattering and temperature calculations, and employs a ray tracing method for imaging and line emission. The calculations can be performed on a variety of grid structures, including a Voronoi tessellation. For its line radiative transfer, \textsc{polaris} makes use of the LAMDA database \citep[Leiden Atomic and Molecular Database;][]{Schoiler2005}, which provides pre-computed molecular parameters, such as quantum numbers, energy levels, Einstein coefficients, and collision rates with H$_2$.

In this work, we perform line radiative transfer, using the large velocity gradient (LVG) approximation to calculate the level populations \citep{Sobolev1957}. This assumes that the velocity difference between two neighbouring grid cells exceeds the thermal broadening of the line, and therefore the locally emitted photons are able to easily escape the grid. Note that \textsc{polaris} imposes an upper limit of 1~pc to the Sobolev length.

\begin{figure}
	\includegraphics[width=\columnwidth]{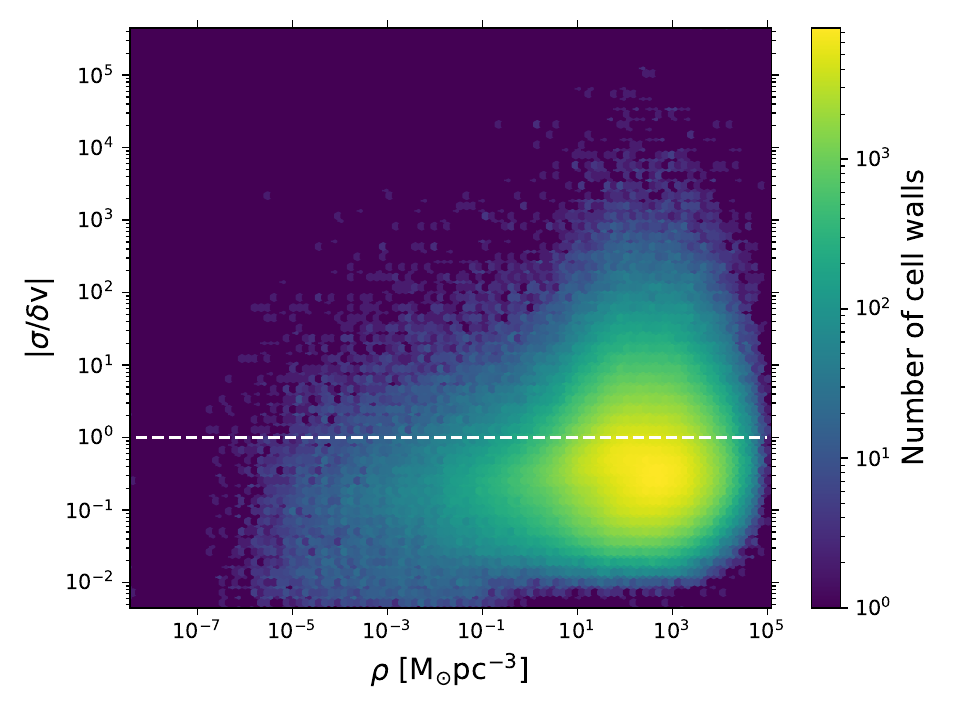}
	\caption{A histogram of interaction length vs cell density (see Equation \ref{eq:interaction-length}). We have assumed that $\sigma = 158$~m/s (HNCO at $T=65$~K). The dashed line denotes an interaction length equal to the separation between two neighbouring cells. Approximately 77\% of the data points lie below the dashed line, which indicates that the LVG approximation is appropriate.}
    \label{fig:interaction-length}
\end{figure}

In order to test the validity of the LVG approximation, we use the photon interaction length, given as \citep{Ossenkopf2002}:
\begin{equation}
    L_{\sigma} = \left |\frac{\sigma}{\mathrm{d}v_n/\mathrm{d}r}\right |,
    \label{eq:interaction-length}
\end{equation}
where $\sigma$ is the thermal broadening of the emission line, and $\mathrm{d}v_n/\mathrm{d}r$ is the local velocity gradient. The LVG approximation should be used when the interaction length for a given cell is smaller than the cell size. In order to check that the condition holds, we consider each Voronoi cell wall, and the two cells that share it. For each of these cells, we compute the velocity component perpendicular to the cell wall. We denote the absolute difference between these two velocity components as $\delta v$. By computing $|\sigma / \delta v|$, we obtain $L_{\sigma}$ in units of the distance between the two cell generating sites. Figure \ref{fig:interaction-length} shows a histogram of $L_{\sigma}$, calculated in this way, as a function of cell density for the TVir snapshot. In this example, we use HNCO and $T = 65$~K, which gives $\sigma \approx 158$~m/s. We can see that the majority of the data points ($77\%$) lie below the dashed line, which marks $L_{\sigma}=1$. Note that the value of $\sigma$ depends on the molecular mass; for the species that we are about to introduce we have $\sigma=146-200$~m/s. This range in $\sigma$ has a very minor effect on the number of datapoints below $L_{\sigma}=1$ ($71-78\%$), which indicates that the LVG approximation remains appropriate for our simulation snapshots.

\begin{table*}
	\centering
	\caption{Summary of molecular tracer data, used as an input for \textsc{polaris}. The abundances have been primarily obtained from the work of \citet{Tanaka2018}, \citet{Riquelme2018} and \citet{Meijerink2011}, and are discussed in more detail in the text. The frequencies of the transitions are taken from the LAMDA database \citep{Schoiler2005}.} 
	\label{tab:abundances}
	\begin{tabular}{lcccr} % four columns, alignment for each
		\hline
		Molecule & Fiducial Abundance & Abundance Range & Transition & Frequency (GHz)\\
		\hline
		CS & $10^{-8}$& $[3 \times10^{-9},3 \times10^{-8}]$ & $J=2-1$ & 97.981\\
		HCN & $10^{-8}$& $[3 \times 10^{-9},3 \times10^{-8}]$	& $J=1-0$ & 88.632\\
		HCO$^+$ & $10^{-8}$& $[3 \times10^{-9},3 \times10^{-8}]$ & $J=1-0$ & 89.189\\
		HNCO & $10^{-8}$& $[3 \times10^{-9},3 \times10^{-8}]$ & $4_{04}-3_{03}$ & 87.925\\
		H$_2$CS & $10^{-9}$& $[3 \times10^{-10},3 \times10^{-9}]$ & $3_{03}-2_{02}$ & 103.040\\
		HC$_3$N & $10^{-8}$& $[3 \times10^{-9},3 \times10^{-8}]$ & $J=10-9$ & 90.979\\
		HNC & $3\times 10^{-9}$& $[10^{-9},10^{-8}]$ & $J=1-0$ & 90.664\\
		N$_2$H$^+$ & $10^{-9}$ & $[3 \times10^{-10},3 \times10^{-9}]$ & $1_{10}-0_{11}$ & 93.172\\
		\hline
	\end{tabular}
\end{table*}

We generate line emission maps of 
CS, HCN, HCO$^+$, HNCO, HNC, H$_2$CS, HC$_3$N and N$_2$H$^+$, which are commonly detected molecules in CMZ clouds \citep[e.g.][]{Jones2012,Rathborne2015,Callanan2021}. The gas temperature is assumed to be 65~K for consistency with the hydrodynamics simulation. This value also matches existing observations of the average gas temperature of the Brick \citep{Ao2013,Ginsburg2016,Krieger2017}. In addition, we adopt a constant abundance for each molecular tracer throughout the cloud. This simplified assumption will likely affect some of the properties of the emission maps, such as the correlation/decorrelation of subregions, the shape of the brightness PDFs and the power spectrum slopes (introduced later in the paper), but it will provide a reasonable starting point of our analysis. 
Indeed, observational studies of the CMZ region, such as that of \citet[see their fig.~18]{Tanaka2018}, show only modest spatial variation of abundances in the plane of the sky. The variation is typically even smaller when we only consider the extend of the Brick cloud. We account for these variations in our modelling by considering three different abundances for each molecular tracer (see below). In addition to abundance variation in the plane of the sky, there can also be variation along the line of sight, which is harder to deduce from observations. However, \citet[see their fig.~5]{Rathborne2014b} found an approximately constant dust-to-emission-line intensity ratio for a few tracers throughout the Brick. For an optically thin tracer (such as HNCO) this constant ratio is consistent with having approximately constant abundance along the line of sight, and it further confirms that our assumptions are sensible. We defer a more in-depth chemical modelling to a future study. The molecular abundances that we assume are obtained primarily from the CMZ observations of \citet[hereafter T18]{Tanaka2018} and \citet[hereafter R18]{Riquelme2018}, and the X-ray dominated region (XDR) models by \citet[hereafter M11]{Meijerink2011}\footnote{The X-ray flux in the CMZ is not particularly large, but the cosmic ray flux is, and the XDR models give a good guide for the behaviour of cosmic-ray-dominated clouds.}. The values for the molecular abundances, as well as the transitions and their frequencies, are summarised in Table~\ref{tab:abundances}. The abundances are chosen as follows: 

\begin{itemize}
    \item \textbf{HCN, HNC:} M11 report abundances of a few times $10^{-9}$ for both species. T18 find $x_{\mathrm{HCN}} \sim 10^{-7}$ and $x_{\mathrm{HNC}} \sim 10^{-8}$, but R18 report $x_{\mathrm{HCN}} \sim 10^{-8}$ for most of their tabulated sources, with $x_{\mathrm{HNC}}$ typically a factor of a few smaller. Therefore, we choose an abundance of $10^{-8}$ for HCN, and $3\times10^{-9}$ for HNC in order to reproduce the ratio of about 3:1 between the integrated intensity of HCN and that of HNC.
    \item \textbf{HCO$^+$:} T18 report abundances of $\sim 10^{-8}$, while R18 report a value a factor of a few smaller.  M11 find that the abundance strongly depends on the choice of cosmic ray ionisation rate. When using cosmic ray ionisation rates comparable to the CMZ, they obtain HCO$^+$ abundances of a few times $10^{-8}$. Therefore, we choose an abundance of $10^{-8}$.
    \item \textbf{N$_2$H$^+$:} T18 report an abundance of a few times $10^{-9}$ for most of the CMZ, and slightly higher in Sgr B2, but R18 give a value of around $10^{-9}$ for most sources. We assume $10^{-9}$ in this work.
    \item \textbf{CS:} T18 find abundances between $10^{-8}$ and a few times $10^{-8}$, while M11 report values of around $10^{-8}$ for cosmic ray ionisation rates consistent with the CMZ. R18 also find values of $\sim 10^{-8}$, and hence we assume an abundance of $10^{-8}$.
    \item \textbf{HC$_3$N:} T18 report an abundance of $\sim 3 \times 10^{-9}$ for most of the CMZ, increasing to $\sim 10^{-8}$ in Sgr B2. R18 also report values closer to $10^{-8}$ in most sources. In consideration of these studies, we assume a fiducial abundance of $10^{-8}$.
    \item \textbf{HNCO:} The abundance of this molecule was not included in T18 or M11, but R18 report values between $10^{-8}$ and a few times $10^{-8}$. \citet{Churchwell1986} find a value of $3 \times 10^{-9}$ in Sgr B2, and hence we assume an abundance of $10^{-8}$.
    \item \textbf{H$_2$CS:} The abundance of this molecule was not reported in T18, M11 or R18. Therefore, we assume a value of $10^{-9}$, to match the observations of local clouds \citep[e.g.][]{Minh1991}.
\end{itemize}

In addition to the abundances listed above, we consider a plausible abundance range for each molecular species (listed in Table \ref{tab:abundances}). We use the extremes of these ranges to create sets of minimum abundance and maximum abundance emission maps alongside our main ones, in order to check the robustness of our results.

Using the above setup, we create a set of emission maps in different velocity channels. The line of sight of the emission maps coincides with that of the real Brick as viewed from Earth. We use 101 velocity channels, covering the radial velocity range between $-50\mathrm{km}~\mathrm{s}^{-1}$ and $50\mathrm{km}~\mathrm{s}^{-1}$, which includes all kinematic parts of the clouds. We have also performed tests with narrower velocity channels, which do not produce significantly different results. Note that the symmetry of the velocity range about 0 is internally imposed by \textsc{polaris}, whereas the velocity distributions of the simulated clouds are biased towards positive values (with a mean of $\sim 20$~km/s, which is consistent with the real Brick; see Appendix \ref{sec:app-moments}). Additionally, we do not include any sub-grid turbulent broadening of the lines. The emission maps are then integrated over the velocity range to create the final \textsc{polaris} output images. These integrated emission maps have dimensions of $3008\times3008$ pixels, and a pixel size of $\sim 0.013 \textrm{pc}$, corresponding to $0.33^{''}$ at the distance of the CMZ. We justify the choice of pixel size in Section~\ref{sec:resolution}.

\subsection{Processing with \textsc{casa}}

\begin{figure*}
	\includegraphics[width=\textwidth]{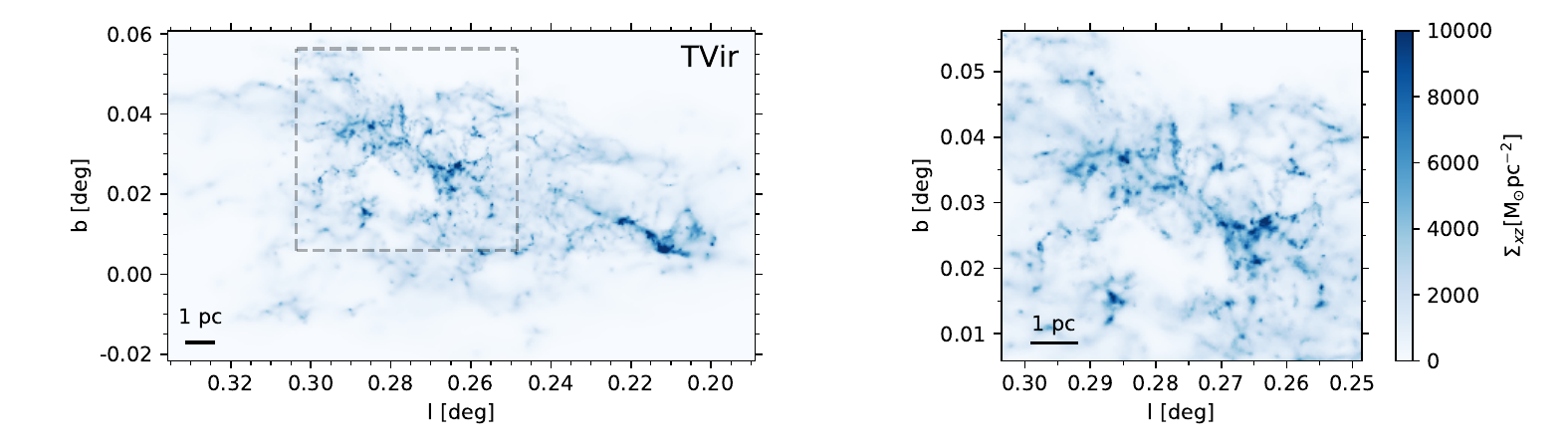}
	\includegraphics[width=\textwidth]{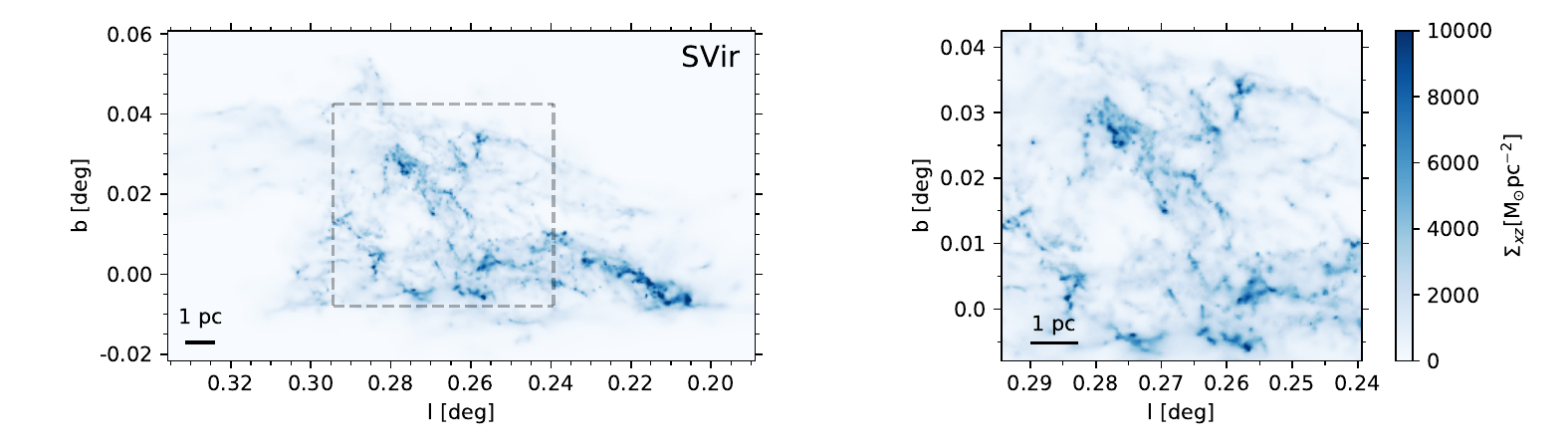}
	\caption{Column density map of the full simulated clouds (\textit{left}), and the high-density regions, selected as the observation areas for the synthetic ALMA emission maps. (\textit{right}). The top panels show the TVir snapshot, and the bottom ones show the SVir snapshot. Note that there is a small offset between the positions of the right-hand panels (also shown as dashed rectangles on the left), which comes from the lopsided mass distribution in each cloud. The colour bars indicate the column densities in each of the panels.}
    \label{fig:column-dens}
\end{figure*}

As the final step towards the creation of synthetic ALMA observations from our simulation snapshots, we process the integrated emission maps from \textsc{polaris} with the \textsc{casa} 5.4.1-31 package \citep[Common Astronomy Software Applications][]{McMullin2007}. Since we are insterested in performing a direct comparison with the data set of \citet{Rathborne2015}, we closely follow their observational input parameters. Therefore, for each molecular tracer we generate six observations using the same array configurations (and hence the same $uv$-coverage) as \citet{Rathborne2015}. Each observation is performed with the \texttt{simobserve} task and it consists of a $72''\times 162''$ mosaic, centred around the densest part of the cloud (see the right-hand panels of Figure~\ref{fig:column-dens}). Each observation is run for 40 minutes on-source, with a precipitable water vapour of ${\rm PWV}=1.5$~mm. We then combine the six observations per emission line and clean the resulting image with the \texttt{tclean} task. The synthesised beam of the final images has an angular size of $\sim 1.9^{''}$ ($\sim0.08$~pc).

\subsection{Resolution}
\label{sec:resolution}

\begin{table*}
	\centering
	\caption{Smallest resolution elements of the SPH snapshots (tidally-virialised and self-virialised), their corresponding regularised Voronoi grids, the chosen \textsc{polaris} output, and the ALMA setup.} 
	\label{tab:resolution}
	\begin{tabular}{lccr} % four columns, alignment for each
		\hline
		SPH smoothing length [pc] & Voronoi cell size [pc] & \textsc{polaris} pixel size [pc] & ALMA beam [pc]\\
		\hline
		$h_{\rm min,Tvir} = 0.014$ & $l_{\rm min,Tvir} = 0.021$ & $d_{\rm pix,Tvir} = 0.013$ & $b_{\rm maj} = 0.080$\\
		$h_{\rm min,Svir} = 0.017$ & $l_{\rm min,Svir} = 0.022$ & $d_{\rm pix,Svir} = 0.013$ & $b_{\rm min} = 0.056$\\
		\hline
	\end{tabular}
\end{table*}

\begin{figure*}
	\includegraphics[width=\textwidth]{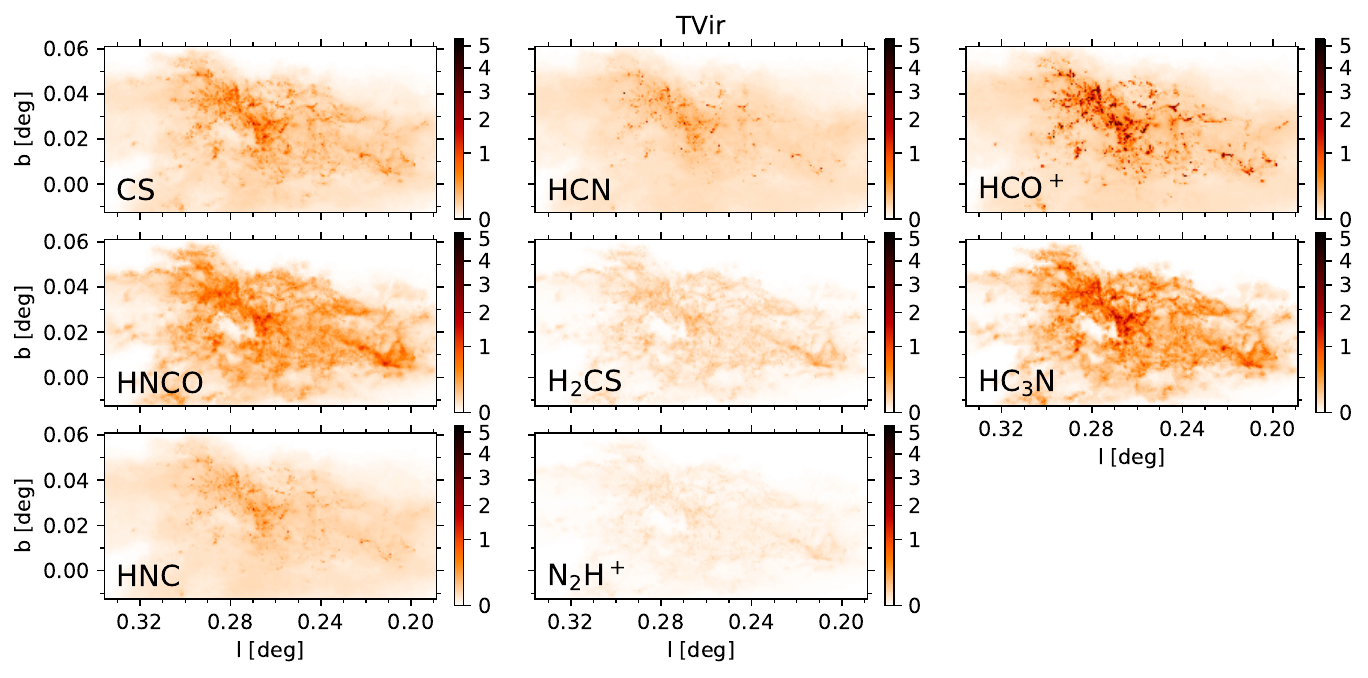}
	\includegraphics[width=\textwidth]{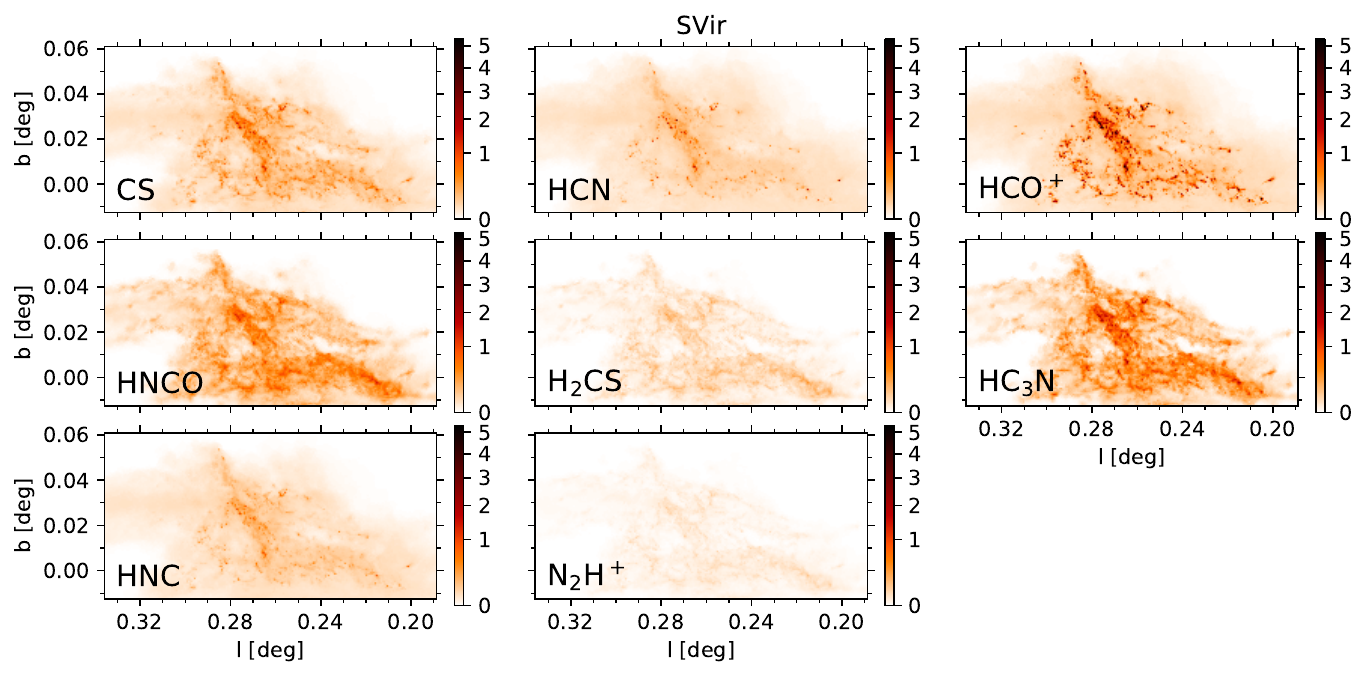}
	\caption{Integrated emission maps of the TVir (\textit{top half of the panels}) and the SVir (\textit{bottom half of the panels}) simulation snapshots, generated with \textsc{polaris}. Each panel contains a label of its corresponding molecular species. The colour bars are in units of mJy/arcsec$^2$ km/s.}
    \label{fig:emission-maps-model}
\end{figure*}

\begin{figure*}
	\includegraphics[width=\textwidth]{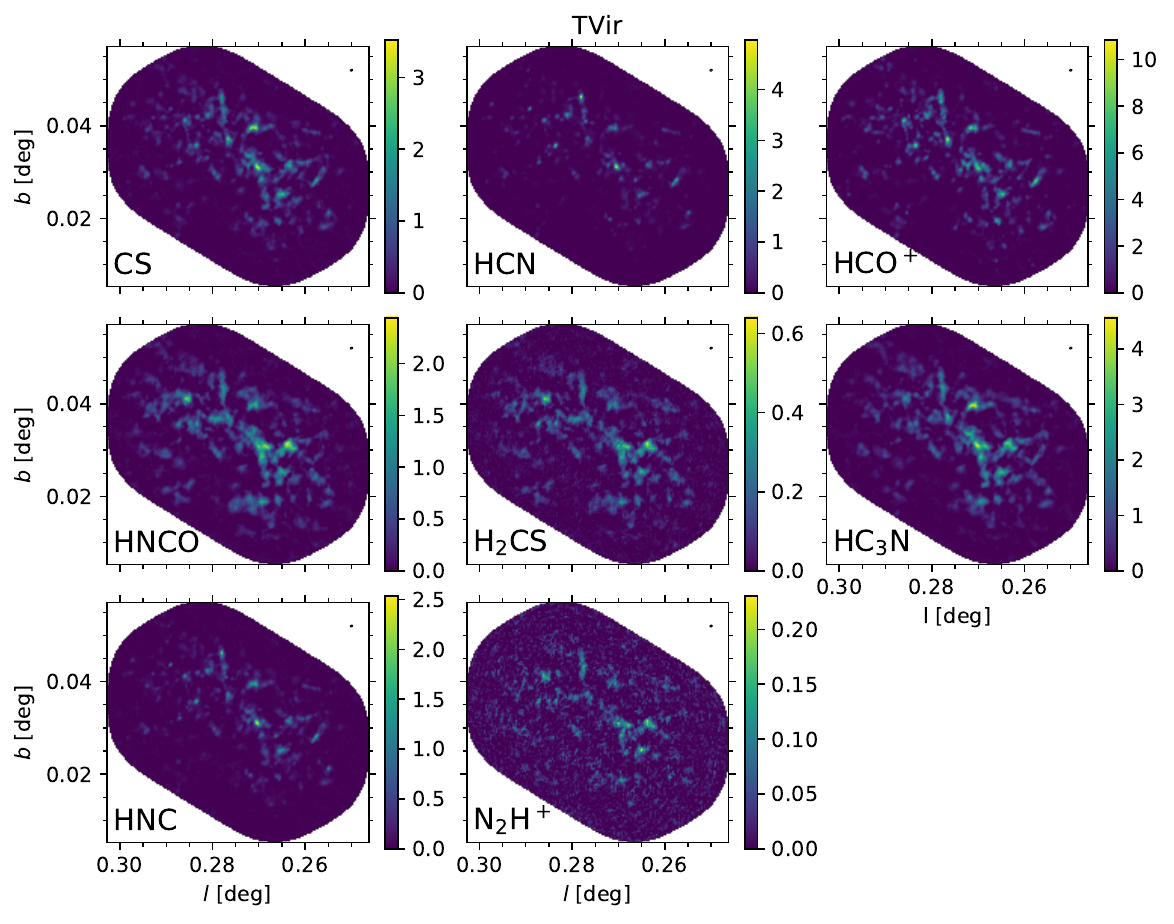}
	\caption{Synthetic integrated emission maps of the central part of the TVir simulation snapshot, generated with \textsc{casa}. The synthesised beam is drawn in the top right corner of each image. The colour bars are in units of mJy/beam km/s.}
    \label{fig:emission-maps-casa}
\end{figure*}

It is crucial to verify whether the finite resolution of the SPH snapshots or of the intermediate post-processing steps can affect the reliability of the final synthetic emission maps. In order to do this, we compare the spatial resolutions of the SPH simulations, the Voronoi grids, the \textsc{polaris} output, and the synthesised beam of the ALMA setup (listed in Table~\ref{tab:resolution}). While the latter two have constant resolution element sizes throughout the cloud, the former two consist of a wide range of particle and cell sizes, respectively. SPH is a Lagrangian method, in which the resolution is defined by a particle mass and not by a spatial scale. Nonetheless, each SPH particle has an associated smoothing length $h$, which is dynamically adjusted based on the local gas density. The densest regions of the clouds consist of the most compact particles, where we measure the minimum smoothing lengths of $h_{\rm min,Tvir} = 0.014~\mathrm{pc}$ (tidally-virialised cloud), and $h_{\rm min,Svir} = 0.017~\mathrm{pc}$ (self-virialised cloud). These high-density regions are likely to have strong contributions to the synthetic emission maps, and the fact that their local resolution (of a few times $h$) is comparable to the synthesised beam size ($b_{\rm maj}=0.080~\textrm{pc}$) suggests that the SPH snapshots resolve the density peaks sufficiently in order to produce robust synthetic emission maps.

However, since the SPH snapshots are not directly post-processed with radiative transfer, but are first mapped onto a grid, we also need to consider the cell sizes. Similarly to the SPH simulations, the Voronoi grids have smaller cell sizes in regions of high density. Here we define the cell size as the cube root of the cell volume, due to the irregular shapes of the grid cells. We find the minimum cell sizes of the two clouds to be $l_{\rm min,Tvir} = 0.021~\textrm{pc}$ and $l_{\rm min,Svir} = 0.022~\textrm{pc}$ (tidally-virialised and self-virialised cloud, respectively). These values are between their corresponding $h_{\rm min}$ and $2h_{\rm min}$, which means that each high-density SPH particle is mapped onto multiple Voronoi cells. This is a desirable property and it indicates some level of resolution preservation.

Finally, we also need to consider the pixel size of the \textsc{polaris} output, which is $d_{\rm pix} = 0.013~\textrm{pc}$. This value is smaller than the minimum Voronoi cell sizes, ensuring that at least one ray passes through each grid cell. However, even if this were not the case, \textsc{polaris} can perform ray splitting to ensure that at least one ray passes through each cell, and hence no cells remain ``invisible'' to the ray-tracing algorithm. Additionally, the \textsc{polaris} pixels need to be smaller than the ALMA beam size. This is true by construction, since we have $d_{\rm pix} < l_{\rm min} < 2h_{\rm min} < b_{\rm min} < b_{\rm maj}$.

The above considerations suggest that the final emission maps should be robust and reliable. We have also tested this by altering the sizes of some of the of the underlying resolution elements, such as
the SPH smoothing lengths and the \textsc{polaris} pixel size. We have created two smoothed (lower resolution) versions of the SPH snapshot by multiplying the smoothing lengths by a factor of 1.5 and 2 respectively. Independently from these tests, we have also created synthetic emission maps using an increased \textsc{polaris} pixel size of 0.04~pc. All of the above tests caused only
minor changes of the final results.

\section{Synthetic emission maps}
\label{sec:comparison}
\begin{figure*}
	\includegraphics[width=\textwidth]{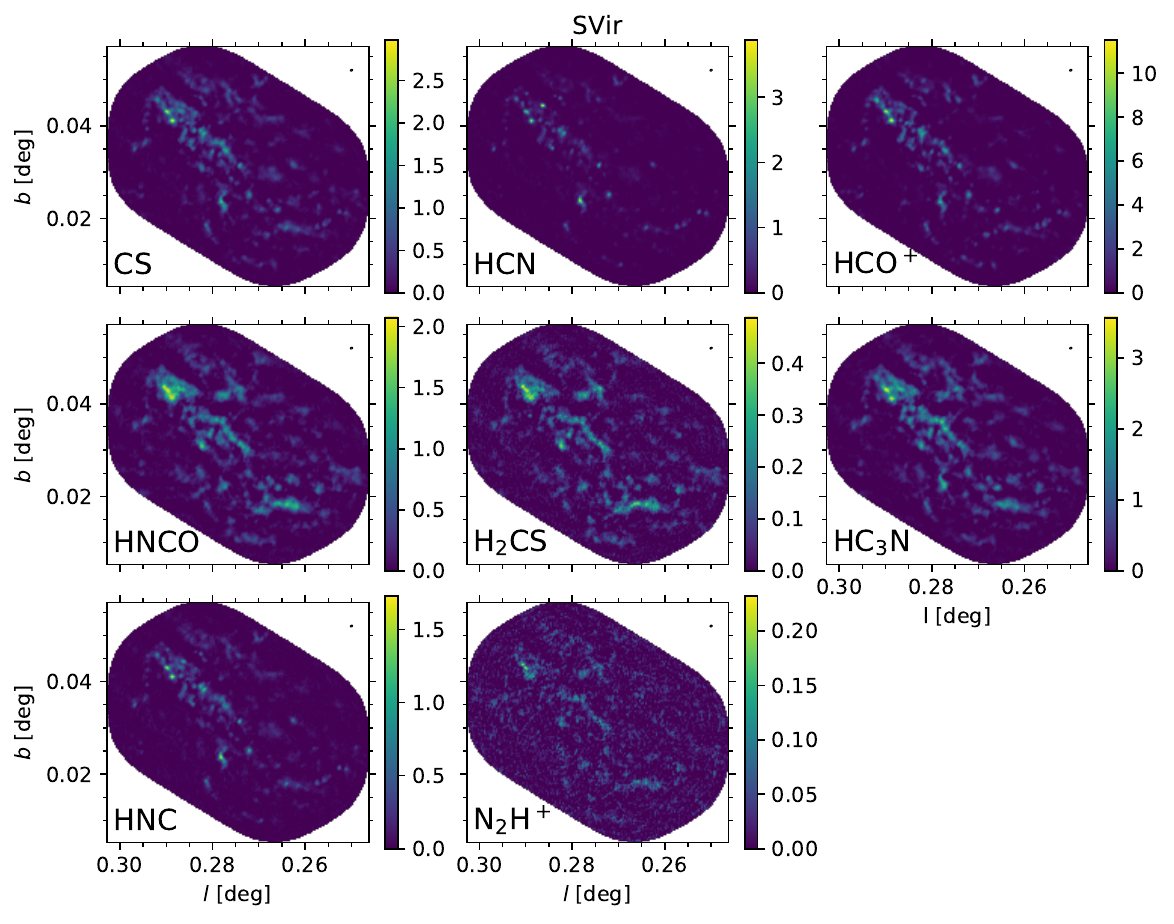}
	\caption{Synthetic integrated emission maps of the central part of the SVir simulation snapshot, generated with \textsc{casa}. The synthesised beam is drawn in the top right corner of each image. The colour bars are in units of mJy/beam km/s.}
    \label{fig:emission-maps-casa-sv}
\end{figure*}

By applying the procedure described in Section~\ref{sec:methods}, we produce integrated emission maps, as well as synthetic ALMA observations of the TVir and SVir simulation snapshots. The integrated emission maps produced with \textsc{polaris} are shown in Figure~\ref{fig:emission-maps-model}. Comparing each of the emission lines, we see that their corresponding maps are visually distinct from one another, with some of them combining diffuse emission with bright small-scale emission peaks (e.g.\ HCN, HCO$^+$, HNC), whereas others trace larger structures (e.g.\ HNCO). The differences in the size scale of the structures result from the fact that each molecule traces different densities within the clouds. In addition to the size scale variations, the emission maps span a range of peak intensities ($0.15-10$~mJy/arcsec$^2$~km/s). If we compare the TVir and the SVir emission maps for a given molecule, we see that they agree with each other in terms of peak intensity and visual appearance. This is not surprising, since the two simulation snapshots contain a similar range of densities (see Figure~\ref{fig:dens-pdfs} and ~\ref{fig:column-dens}). 

\begin{table*}
	\centering
	\caption{Maximum peak intensities of the synthetic ALMA observations of the TVir and SVir simulation snapshots, and comparison to the peak intensities reported by \citet{Rathborne2015}. For each simulation snapshot, we give the peak intensities of the emission maps with the fiducial abundances, as well as those with the minimum and maximum plausible abundances (see Table~\ref{tab:abundances}). Additionally, we provide the peak intensities both in units of [mJ/beam km/s] and [K km/s].} 
	\label{tab:peaks}
	\begin{tabular}{lcccccccccc} 
		\hline
		 Molecule & \multicolumn{5}{|c|}{Maximum peak intensity [mJy/beam km/s]}& \multicolumn{5}{|c|}{Maximum peak intensity [K km/s]}\\
		 & TVir &[min,max] & SVir &[min,max] & Rathborne+15 & TVir &[min,max] & SVir &[min,max] & Rathborne+15\\
		\hline
		CS & 3.5 &[2.2, 4.8] & 3.0 & [1.7, 4.8]& -- & 0.17 & [0.11, 0.23] & 0.15 & [0.10, 0.24]& --\\
		HCN & 5.0 &[2.4, 12.2] & 3.9 & [1.6, 12.2]& 5.3 & 0.30 & [0.14, 0.73] & 0.23 & [0.10, 0.50]& 0.32\\
		HCO$^+$ & 10.8 &[5.5, 21.8] & 11.5 & [5.7, 21.8]& 4.2 & 0.64 & [0.32, 1.29] & 0.68 & [0.34, 1.48]& 0.25\\
		HNCO & 2.4 &[1.1, 4.1] & 2.1 & [0.80, 4.1]& 4.2 & 0.15 & [0.07, 0.25] & 0.13 & [0.05, 0.23]& 0.25\\
		H$_2$CS & 0.64 &[0.19, 1.5] & 0.49 & [0.20, 1.5]& 0.5 & 0.03 & [0.01, 0.07] & 0.02 & [0.01, 0.05]& 0.02\\
		HC$_3$N & 4.6 &[2.6, 6.7] & 3.6 & [2.1, 6.7]& -- & 0.26 & [0.15, 0.38] & 0.20 & [0.12, 0.28]& --\\
		HNC & 2.5 &[1.5, 5.0] & 1.7 & [1.0, 5.0]& -- & 0.14 & [0.09, 0.29] & 0.10 & [0.06, 0.21]& --\\
		N$_2$H$^+$ & 0.23 &[0.12, 0.52] & 0.23 & [0.16, 0.52]& -- & 0.012 & [0.006, 0.03] & 0.012 & [0.009, 0.02]& --\\
		\hline
	\end{tabular}
\end{table*}

Figure~\ref{fig:emission-maps-casa} and \ref{fig:emission-maps-casa-sv} show the cleaned synthetic ALMA observations of the TVir and SVir snapshots, respectively. Similarly to the \textsc{polaris} emission maps from Figure~\ref{fig:emission-maps-model}, we see some differences between the ALMA images of the different molecular species, in terms of the cloud structures that they trace. However, relative to Figure~\ref{fig:emission-maps-model}, the synthetic observations are affected by the procedure of making an interferometric observation. For instance, the large-scale, diffuse emission that is well traced by HNC and HCO$^+$ in Figure~\ref{fig:emission-maps-model}, cannot be seen in Figure~\ref{fig:emission-maps-casa} and \ref{fig:emission-maps-casa-sv} due to the limited uv-coverage of the synthesised beam. Through visual inspection, we find that the relative brightness of peaks in the synthetic ALMA observations vary between the different molecular tracers. Furthermore, as in Figure~\ref{fig:emission-maps-model}, the maximum peak intensity also varies with the choice of molecular tracer. Table~\ref{tab:peaks} provides a summary of the peak intensities from Figures~\ref{fig:emission-maps-casa} and \ref{fig:emission-maps-casa-sv}, and also provides reference values from the observations of \citet{Rathborne2015}. We see that the TVir and SVir snapshots produce ALMA emission maps with very similar maximum peak brightness. Additionally, there is good agreement between the values for HCN, HCO$^+$, HNCO and H$_2$CS obtained from observations, compared to the synthetic ALMA observations of the two simulation snapshots. Note that the emission maps for HCN, HCO$^+$ and HNCO presented in \citet{Rathborne2015} combine the ALMA observations with additional single dish data, which slightly boosts the intensity peaks. The synthetic HCO$^+$ emission maps contain brighter peaks than the real Brick observations, which may be due to overestimation of the molecular abundance in our modelling.

\begin{figure*}
	\includegraphics[width=\textwidth]{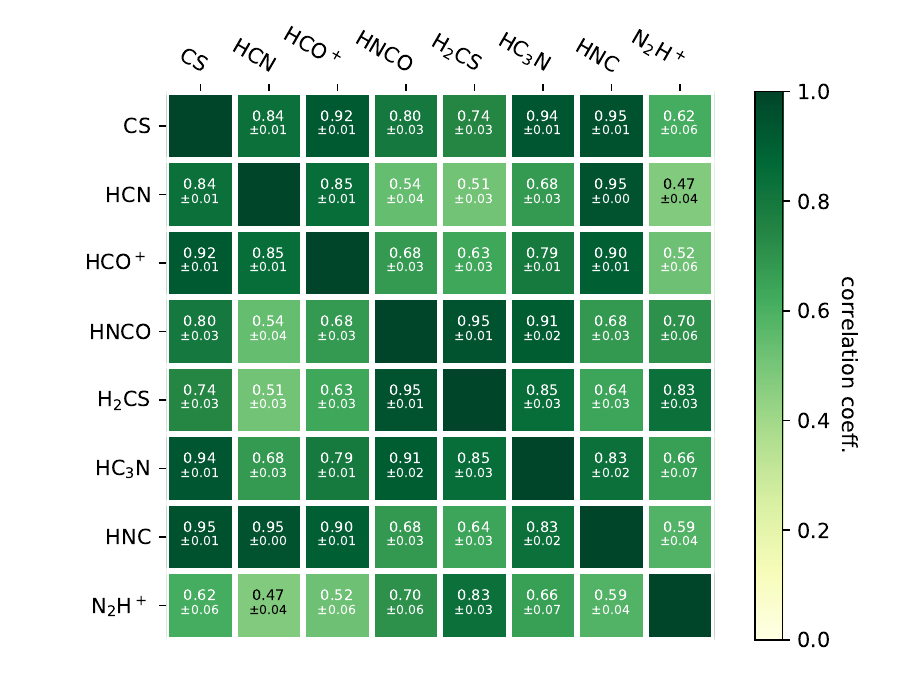}
    \caption{Cross-correlation coefficients of each molecular tracer with respect to each of the others for the TVir snapshot. The uncertainties of all coefficients are estimated to be $<0.1$ by using a bootstrap technique.}
    \label{fig:corr-matrix}
\end{figure*}

\begin{figure*}
	\includegraphics[width=\textwidth]{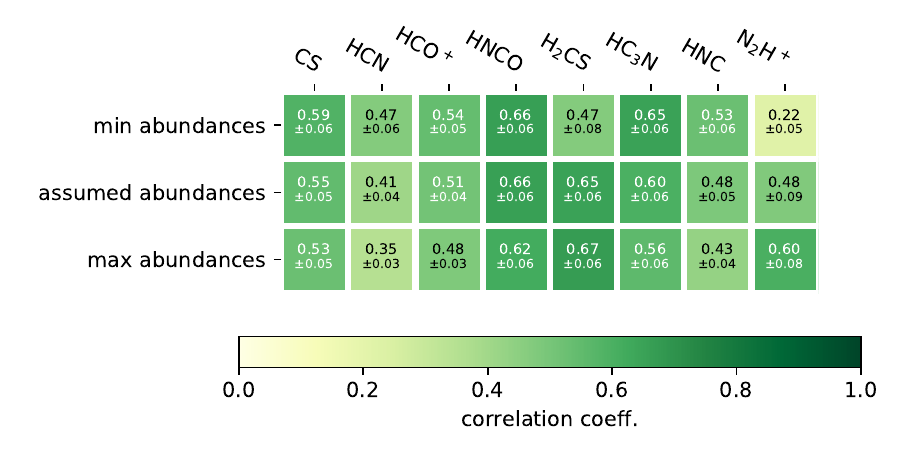}
    \caption{Cross-correlation coefficients of each molecular tracer with respect to the column density of the TVir snapshot. The three rows correspond to the chosen abundances (\textit{middle}) and their minimum (\textit{top}) and maximum (\textit{bottom}) range of trusted values (see Table~\ref{tab:abundances}). The uncertainties of all coefficients are estimated to be $<0.1$ by using a bootstrap technique.}
    \label{fig:corr-dens}
\end{figure*}

In addition to a simple visual inspection, we can quantify the differences in the synthetic ALMA emission maps by computing the cross-correlation coefficients between them. Figure~\ref{fig:corr-matrix} shows the cross-correlation coefficients of the TVir emission maps. The results for the SVir emission maps are omitted due to their similarity to the TVir results. We find that all of the molecular lines are well correlated, with coefficients in the range $0.47-0.95$. We estimate the uncertainties of these coefficients to be $<0.1$ using a bootstrap technique in which we calculate the coefficients for many sub-regions of the full emission maps. In Figure~\ref{fig:corr-matrix} we can identify two groups of molecular tracers, which have strong correlations within the group, and have only moderate correlations outside the group. The first group consist of CS, HCN, HCO$^+$ and HNC, while the second group consists of HNCO, H$_2$CS, HC$_3$N and N$_2$H$^+$. Upon visual inspection, we find that the latter group traces some of the larger structures, while the former group mainly traces emission from compact peaks. It is worth noting that while fitting in with the latter group, N$_2$H$^+$ is somewhat of an outlier. This is because the N$_2$H$^+$ emission maps are largely dominated by noise, due to their overall low brightness.

If we consider only the molecular species which were also observed by \citet{Rathborne2015} (i.e.\ HCN, HCO$^+$, HNCO and H$_2$CS), the range of cross-correlation coefficients that we calculate becomes $0.51-0.95$. The observations of \citet{Rathborne2015} have a broader range of cross-correlation coefficients (and somewhat lower, with $0.28-0.81$). Encouragingly, the coefficients are distributed in the same way between the molecular couples, i.e.\ the lower/higher cross-correlation coefficients in the simulated emission maps correspond to the lower/higher cross-correlation coefficients in the observations. Specifically, HCN and HCO$^+$ are most strongly correlated with each other, as are HNCO and H$_2$CS. This reflects the difference in excitation energy between both pairs of lines (see Table~1 of \citealt{Rathborne2015}).

\citet{Rathborne2015} also compute the cross-correlation coefficients between the molecular line emission maps and the dust continuum emission map, the latter of which is expected to trace the gas density. They find that none of their molecular emission maps are strongly correlated with the continuum and hence the gas density, with cross-correlation coefficients in the range $0.09-0.39$. In order to compare to their results, we compute the cross-correlation coefficients between the gas column density and the synthetic emission maps (see Figure~\ref{fig:corr-dens}). In reality, the continuum emission does not perfectly trace the column density, since there are temperature fluctuations within the Brick. However, at 3~mm we probe the Rayleigh-Jeans tail of the dust emission, which is only weakly sensitive to temperature. Under these circumstances, using the column density and the dust continuum emission interchangeably is a reasonable approximation. Figure~\ref{fig:corr-dens} shows the cross-correlation coefficients between the column density map of the TVir snapshot and the molecular emission maps from Figure~\ref{fig:emission-maps-casa}, which we find to be in the range $0.41-0.66$ for the assumed abundances. In addition, we have reconstructed the maps in Figure~\ref{fig:emission-maps-casa} for the minimum and maximum plausible abundances (see Table~\ref{tab:abundances}). With the exception of N$_2$H$^+$, we do not see a strong variation of the cross-correlation coefficients with the molecular abundance. We find that the best tracer of the column density is HNCO, which is in agreement with the findings of \citet{Rathborne2015}, even though they report a significantly weaker correlation. The fact that our synthetic observations correlate a lot more strongly with the density structure than the real Brick observations is likely due to one or more of the simplifying assumptions that we have made. In particular, we have assumed constant temperature and abundances throughout the simulated clouds, while the real Brick has variations in temperature and abundances \citep[e.g.][]{Mills2015}. Our models provide a first attempt at reproducing the different emission lines of the Brick, and can be improved in future work by including cooling, heating, chemistry, and more detailed modelling of the molecular species.

\section{Cloud structure}
\label{sec:cloud-structure}
Similarly to the emission maps of \citet{Rathborne2015}, our synthetic maps show complex, hierarchical morphology. In this section, we aim to quantify the structure of the simulated Brick, using a variety of quantitative metrics, such as density probability distribution functions (PDFs), the fractal dimension, the spatial power spectrum, and the moments of inertia (through the J-plots method; \citealt{Jaffa2018}). Each of these metrics probes a different substructural property of the synthetic emission maps. The density PDFs show the relative occurrence rate of pixels of different brightness, while the power spectrum is sensitive to the spatial separation of the brightness peaks. The fractal dimension is a way of quantifying complexity in substructures of the cloud. A high fractal dimension indicates a turbulent cloud with winding contours. The opposite of that would be a low fractal dimension cloud, where each contour is close to circular, and it likely encloses a gravity-dominated region. Finally, the moments of inertia measured through the J-plots method characterises the shapes of the cloud substructures, and it divides them into centrally condensed, centrally rarefied, and elongated/filamentary structures.

\begin{figure}
	\includegraphics[width=\linewidth]{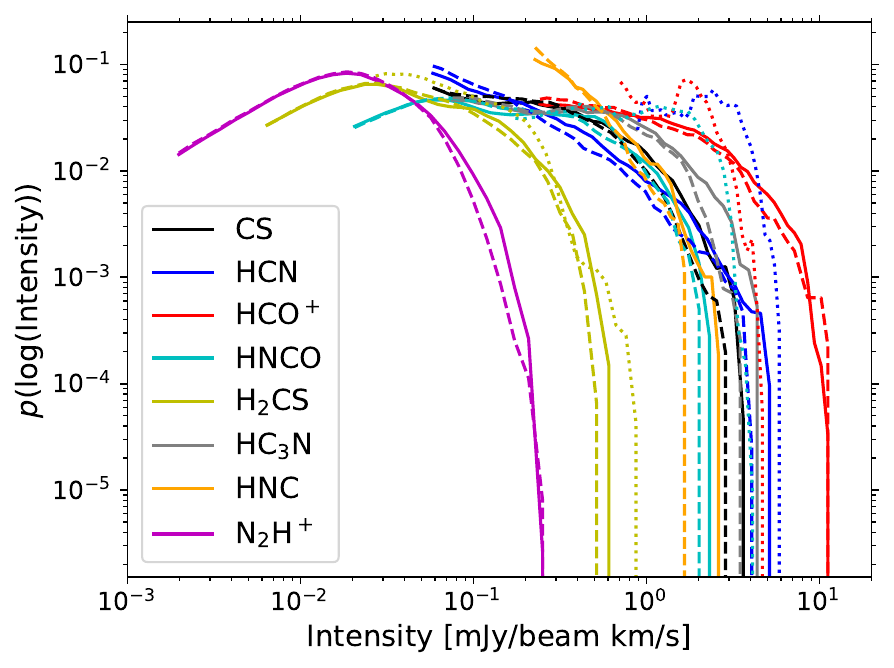}
	\caption{Brightness PDFs of the synthetic and real ALMA observations. The solid and dashed lines show the TVir and SVir emission maps respectively (from Figure~\ref{fig:emission-maps-casa} and~\ref{fig:emission-maps-casa-sv}), and the four dotted lines show the observations of HCN, HCO$^+$, HNCO and H$_2$CS from \citet{Rathborne2015}.}
    \label{fig:pdfs-brighness}
\end{figure}

\begin{figure}
	\includegraphics[width=\linewidth]{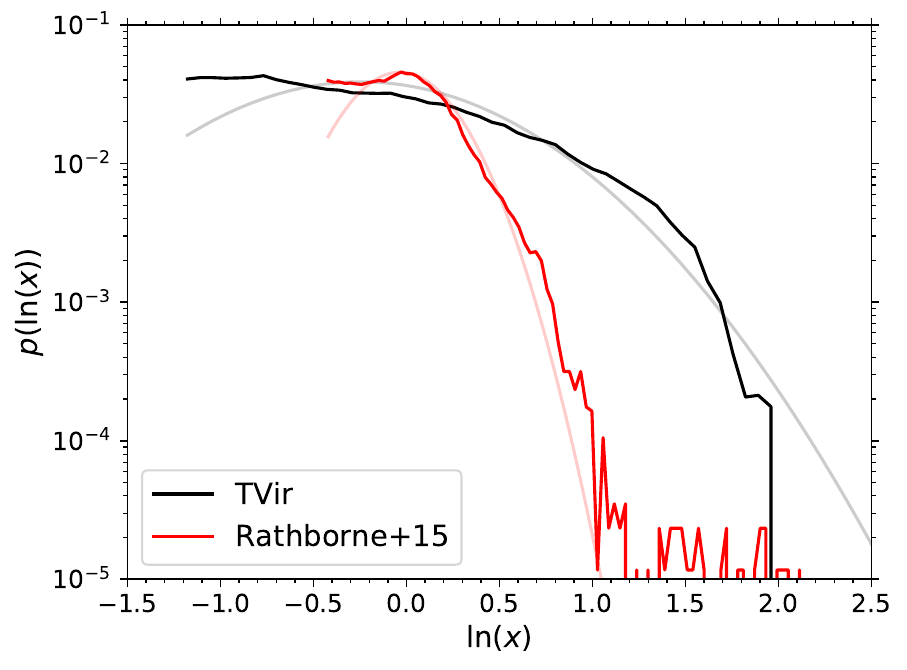}
	\caption{Density PDFs of the 3~mm continuum emission in the simulated \citep{Kruijssen2019} and real Brick \citep{Rathborne2015}. The $x$-axis is given in terms of the natural logarithm of the pixel intensity divided by the mean pixel intensity ($\ln(x)$). The semi-transparent lines are log-normal fits to the PDFs of the the corresponding colour.}
    \label{fig:pdfs-intensity}
\end{figure}

\subsection{Brightness and density probability distribution functions}
\label{sec:pdfs}

We construct brightness PDFs of the TVir and SVir \textsc{casa} emission maps, as well as of the \citet{Rathborne2015} data, in order to compare the relative occurrence of pixels of different brightness (see Figure~\ref{fig:pdfs-brighness}). Since it is more physically meaningful to consider column density or volume density PDFs, and continuum emission traces the density structure better than line emission, we also include the PDFs of the 3~mm continuum emission from \citet{Kruijssen2019} (using the TVir snapshot only) and \citet{Rathborne2015} in Figure~\ref{fig:pdfs-intensity}. Note that the $x$-axis of Figure~\ref{fig:pdfs-intensity} is given in terms of $\ln(x)$, where $x$ is the ratio of a given pixel value and the mean pixel value included in the PDF. In the case of the continuum PDFs, we assume that the gas column density is proportional to the dust column density, which in turn is proportional the pixel intensity, and hence $x$ also equals the ratio between the column density and the mean column density. Additionally, we have excluded pixels under a certain threshold
(which differs between the images)
%(HNCO: 0.1~mJ/beam~km/s (TVir), 0.62~mJ/beam~km/s (observation); 3~mm continuum: 0.17~mJ/beam (TVir), 0.33~mJ/beam (observation)), 
in order to separate the background noise from the emission, and to ensure that the PDFs only contain contributions from closed contours within the images \citep{Alves2017}. 

Figure~\ref{fig:pdfs-brighness} shows good agreement between the PDFs of the TVir and SVir simulations (solid and dashed lines respectively), which can be explained by similarities in the underlying density distributions of the two snapshots (see Figure~\ref{fig:dens-pdfs} and~\ref{fig:column-dens}).
The shapes of the PDFs are approximately log-normal with a flattening at lower intensities. We also include the brightness PDFs of the molecular species that are observed by \citet{Rathborne2015} for comparison (i.e. HCN, HCO$^+$, HNCO and H$_2$CS; dotted lines). We find good agreement between the simulated and observed brightness distributions for H$_2$CS. The other molecules show some discrepancies in terms of the peak brightness (as already shown in Table~\ref{tab:peaks}) and the slope of the higher-brightness end of the curves. Many of these discrepancies can be overcome by tuning the assumed abundances, and by more generally improving the chemical modelling. In particular, by slightly increasing the abundance of HNCO and slightly decreasing that of HCO$^+$, we can mostly compensate for the discrepancies between the simulated and observed curves. In contrast, the peak brightness of HCN in the synthetic images is close to that of the real Brick, but there is a mismatch of the high brightness slope that likely requires better chemical modelling. Alternatively, some of these discrepancies may be arising from differences in the gas structure between the simulations and the Brick.
%Moreover, both curves are well-described by log-normal distributions of similar width ($\sigma_{\ln x,{\rm TVir}}=0.56$; $\sigma_{\ln x,{\rm SVir}} = 0.59$). The PDF of the \citet{Rathborne2015} line emission map is also approximately log-normal in shape, but it is much narrower ($\sigma_{\ln x,{\rm R15}}=0.33$). 

We find similar log-normal shapes in the 3~mm continuum emission (Figure~\ref{fig:pdfs-intensity}), with widths of $\sigma_{\ln x,{\rm TVir}}=0.70$, and $\sigma_{\ln x,{\rm R15}}=0.26$. Note that \citet{Rathborne2014} report a PDF width of $0.34\pm 0.02$ for the same observed 3~mm continuum emission map. The difference here likely arises from a slightly different choice of fitting range for the log-normal curve. In order to consolidate the two results, we will use $\sigma_{\ln x,{\rm R15}}=0.30\pm 0.05$ (and also $\sigma_{\ln x,{\rm TVir}}=0.70\pm 0.05$) in all calculations below. Finally, the observed 3~mm continuum emission map has a high-density tail in its PDF, which deviates from the log-normal shape. This feature is caused by a few bright peaks, which have been attributed to regions of ongoing star formation \citep{Rathborne2014,Walker2021}. The feature is not present in the continuum PDF of the TVir snapshot, despite the more prominent gravitational collapse in the simulation. 
This is due to the fact that the smallest (sub-sink radius; $\rho \geq 10^{-17} \mathrm{g~cm}^{-3}$) scales of star formation remain unresolved in our models, and any density enhancements are rapidly converted into sink particles, which are not included in the emission maps.
%This is due to the creation of sink particles from bound gas with $\rho \geq 10^{-17} \mathrm{g~cm}^{-3}$, which are not included in the emission maps.

There is an enormous body of literature showing that the width of a log-normal \textit{volume} (not column) density PDF can be expressed in terms of a small number of parameters \citep[e.g.][]{Vazquezsemadeni1994,Padoan1997,Krumholz2005,Hennebelle2011,Federrath2012,Molina2012}. These are the turbulence driving parameter, $b$\footnote{Solenoidal turbulence: $b=1/3$; natural mix: $b\approx 0.4$; compressive turbulence: $b=1$; see e.g.\ \citealt{Federrath2016}.}, the 3D sonic Mach number, $\mathcal{M}$, and the turbulent plasma beta, $\beta$. Therefore, in order to link $\sigma_{\ln x,{\rm TVir}}$ and $\sigma_{\ln x,{\rm R15}}$ to the above physical parameters, we first need to obtain the volume density dispersions. To do so, we follow the procedure used in \citet{Federrath2016}, who analyse the PDF of the 3~mm continuum emission map of the Brick, presented in \citet{Rathborne2014}. First, we find the column density dispersion $\sigma_{x} = [\exp(\sigma_{\ln x}^2) -1]^{1/2}$ \citep[e.g.][]{Price2011}, which gives $\sigma_{x,{\rm TVir}} = 0.80 \pm 0.07$ and $\sigma_{x,{\rm R15}} = 0.31\pm 0.05$. Next, \citet{Federrath2016} find the conversion factor between the column density and the volume density dispersion, $\mathcal{R}^{1/2} = 0.28\pm 0.11$, using the method of \citet{Brunt2010}. Note that the large uncertainty on the value of $\mathcal{R}^{1/2}$ comes from the fact that the method assumes an isotropic mass distribution in the cloud, whereas \citet{Federrath2016} find a moderate level of anisotropy in the Brick. They attribute the anisotropy to the cloud's strong ordered magnetic field \citep{Pillai2015}, which is not present in the simulations, and hence we do not expect a larger uncertainty when applying this method to the TVir snapshot. By dividing $\sigma_x$ by $\mathcal{R}^{1/2}$, we get the volume density dispersions $\sigma_{\rho/\rho_0,{\rm TVir}} = 2.8 \pm 1.1$ and $\sigma_{\rho/\rho_0,{\rm R15}} = 1.1\pm 0.5$, where $\rho_0$ is the average volume density. Our estimate for $\sigma_{\rho/\rho_0,{\rm R15}}$ is consistent with that of \citet{Federrath2016}, who find the value $1.3\pm 0.5$.

For the real Brick, \citet{Federrath2016} find that $\mathcal{M}=11 \pm 3$ and $\beta = 0.34 \pm 0.35$, which let them deduce that $b=0.22\pm 0.12$, using the relation 
\begin{equation}
\sigma_{\rho/\rho_0} = b\mathcal{M}(1+\beta^{-1})^{-1/2}.
\label{eq:sig_rho}
\end{equation}
Using our value for $\sigma_{\rho/\rho_0,{\rm R15}}$, we find $b=0.20\pm 0.13$, which is consistent with the above result and with solenoidal turbulence driving. 

Within the simulations, the parameters on the right-hand side of eq.~\ref{eq:sig_rho} are either known or can be measured. First, the simulations do not include a magnetic field, which means that $\beta \rightarrow \infty$, and hence $(1+\beta^{-1})^{-1/2}=1$. Second, by using a Helmholtz decomposition, we can represent the velocity field of the TVir simulation as the sum of a solenoidal component and a compressive component. Using the kinetic energy of the two components, we deduce that the turbulence driving is solenoidal (caused by by the strong shear arising from the Galactic potential), and hence $b\approx0.3$ (Petkova et al. in prep.). Finally, we calculate $\mathcal{M} = 7.7 \pm 0.6$, using the size-linewidth relation of the simulated cloud (Petkova et al. in prep.). Combining these parameters, we find that the right-hand side of eq.~\ref{eq:sig_rho} equals $2.2 \pm 0.2$. This is in agreement with the measurement of $\sigma_{\rho/\rho_0,{\rm TVir}}=2.8\pm1.1$, as expected.

The direct measurements of the above parameters in the simulations allow us to investigate what causes the discrepancy between $\sigma_{\rho/\rho_0,{\rm TVir}}$ and $\sigma_{\rho/\rho_0,{\rm R15}}$, by exploring possible variations in $b$, $\mathcal{M}$, and $\beta$. First, the turbulence driving parameter $b$ is in agreement between the simulated and observed cloud, and it is unlikely to change unless we alter the gravitational potential in the simulation or add other sources of turbulence that cause cloud-scale compression. Second, the Mach number that we measure in the simulated cloud is lower by about $1\sigma$ compared to the observed Brick. If we used the observed value of $\mathcal{M}$, the right-hand side of eq.~\ref{eq:sig_rho} would become even larger and shift away from the $\sigma_{\rho/\rho_0,{\rm R15}}$ value. Instead, the simulation would need to have an even smaller Mach number (by a factor of $\sim 2$) in order to match the observed density PDF. By necessity, this would lower the velocity dispersion of the simulated cloud, and it would cause it to undergo rapid collapse and star formation. Finally, if we were to modify the simulation to include the same turbulent magnetic field strength as the real Brick, the density dispersion would be modified by a factor of $(1+\beta^{-1})^{-1/2} \approx 0.5$. The resulting right-hand side of eq.~\ref{eq:sig_rho} would then predict $\sigma_{\rho/\rho_0}=1.1$, which is in excellent agreement with $\sigma_{\rho/\rho_0,{\rm R15}}=1.1\pm0.5$. The caveat here is that the measured value of $\beta$ in the Brick has a large error bar, and also that by introducing a magnetic field in the simulations, one might also end up altering the Mach number. Despite these concerns, our analysis indicates that the discrepancy in the widths of the density PDFs may be attributed to the lack of a magnetic field in the simulations.

\subsection{Spatial power spectrum}
\label{sec:power-spectrum}
There is no explicit turbulence driving in the simulations of \citet{Dale2019}. Despite that, \citet{Kruijssen2019} find that the clouds maintain large velocity dispersion along their orbit, which is attributed to the presence of strong shear. To study this behaviour we compute the spatial power spectra of the synthetic emission maps, using TurbuStat \citep{Koch2019}. The power spectra are constructed by performing a 2D Fourier transform of the emission maps, and plotting the 1D radial profile of the resulting images as a function of length scale.

\begin{figure}
	\includegraphics[width=\columnwidth]{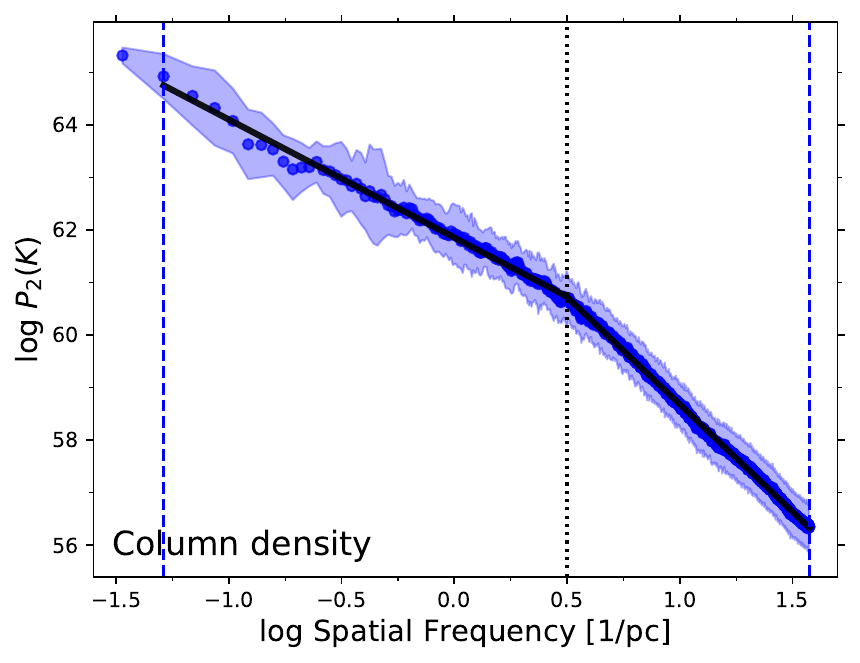}
	\includegraphics[width=\columnwidth]{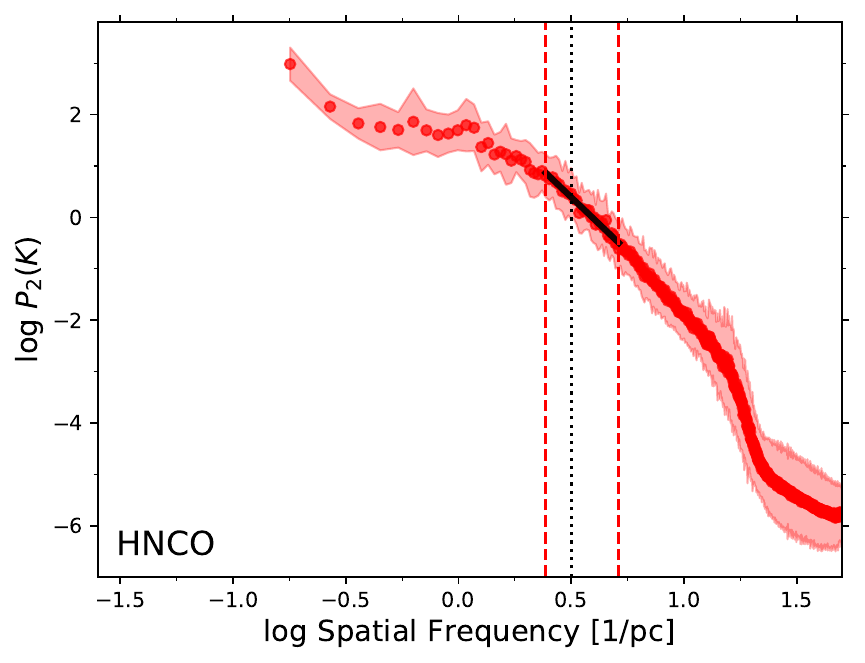}
	\caption{Power spectrum of the column density (top) and the HNCO emission map of the TVir snapshot (bottom). The power spectrum of the column density exhibits a break at 0.32~pc, while the power spectrum of HNCO emission map can be modelled as a single power law within the spatially reliable range. The range of spatial scales used for the line fitting is shown by the vertical dashed lines. The black dotted line marks the position of the break in the power spectrum slope in the column density map.}
    \label{fig:power-spectrum}
\end{figure}

Figure~\ref{fig:power-spectrum} shows the power spectrum of the TVir column density (top panel), alongside the power spectrum of the HNCO emission map of the TVir snapshot (bottom panel). The power spectrum of the column density follows a broken power-law, with a clearly defined break at $\sim 0.32$~pc. Note that the position of the break does not change if we vary the pixel size of the column density map that we use. In contrast, the HNCO emission map power spectrum has a more complex shape, which does not overall resemble a single power law. As a result of the interferometric nature of the synthetic emission maps, the large spatial scales are filtered out, whereas the small spatial scales are correlated due to the beam profile. Therefore, there is a narrow window of spatial scales where we can trust the power spectrum values. We exclude spatial scales below three beam sizes (using the geometric mean of the major and minor beam axes; see \citealt{Koch2020}), which corresponds to values above 5.16~pc$^{-1}$ in the power spectrum. We also exclude spatial scales above the maximum resolvable scale of $10.5^{''}$ (see ALMA Cycle 0, Extended Configuration, Band 3), corresponding to values below 2.38~pc$^{-1}$. This range is marked with vertical dashed lines in Figure~\ref{fig:power-spectrum}, and we find that within it the power spectrum can be fitted with a single power law. The slopes of the remaining TVir and SVir emission maps are summarised in Table~\ref{tab:power-spec}.

\begin{table}
	\centering
	\caption{Power spectrum slopes.}
	\label{tab:power-spec}
	\begin{tabular}{lcc} % four columns, alignment for each
		\hline
		Molecular & \multicolumn{2}{|c|}{Power spectrum slope}\\
		species & (TVir) & (SVir)\\
		\hline
		CS & $-3.7\pm 0.1$ & $-3.5\pm 0.1$\\
		HCN & $-2.7\pm 0.1$ & $-2.3\pm 0.1$\\
		HCO$^+$ & $-3.2\pm 0.1$ & $-3.1\pm 0.2$\\
		HNCO& $-4.2\pm 0.1$ & $-4.0\pm 0.1$\\
		H$_2$CS & $-3.5\pm 0.1$ & $-3.1\pm 0.1$\\
		HC$_3$N & $-4.1\pm 0.1$ & $-4.1\pm 0.1$\\
		HNC & $-3.2\pm 0.1$ & $-2.9\pm 0.1$\\
		N$_2$H$^+$ & $-1.8\pm 0.1$ & $-1.5\pm 0.1$\\
		\hline
	\end{tabular}
\end{table}

The power spectrum of the TVir column density has two linear slopes: $\alpha_1=-2.25 \pm 0.01$ for large spatial scales, and $\alpha_2=-4.09 \pm 0.01$ for small spatial scales. This is similar to the results of \citet{Rathborne2015}, who find a break in the 3~mm continuum emission power spectrum of the real Brick at 0.12~pc, with $\alpha_1 \sim -2.5$ and $\alpha_2 \sim -8.0$. However, note that the break in the power spectrum of the real Brick occurs very close to the beam scale, which means that it is possible for it to be an observational artefact \citep[see][]{Koch2020}. The differences in the power spectrum parameters between the real Brick and the simulation reflect differences in the underlying physical parameters. For example, one explanation for the slight discrepancy in $\alpha_1$ may be the fact that the simulations do not include the effects of magnetic fields. Indeed, \citet{Padoan2004} find that their simulation with equipartition of kinetic and magnetic energy produces a slope of $-2.25 \pm 0.01$, while their super-Alfv\'enic one generates a steeper slope of $-2.71 \pm 0.01$. However, there are many factors that influence the slope of the power spectrum, and the responsible physical processes cannot be uniquely identified. 

Typically, a break in the power spectrum occurs at a size scale where energy is being injected into the system, or dissipated more efficiently. Since the effect of the shear (and hence turbulent energy injection) is negligible at this size scale, we consider energy dissipation due to thermal processes or gravitational collapse. 
A potentially relevant length scale for these effects is the Jeans length. The average volume density within the simulations is difficult to define due to their prevalent substructure, but sensible values are bracketed by the initial volume-averaged volume density and the evolved mass-weighted average volume density for them is in the range $0.4{-}6 \times 10^4~\mathrm{cm}^{-3}$. This corresponds to a Jeans length of $0.13{-}0.50~{\rm~pc}$.
%The relevant length scales for these effects are the sonic scale and Jeans length, respectively. The average mass-weighted volume density of the TVir snapshot is $\approx 6 \times 10^4~\mathrm{cm}^{-3}$, which corresponds to a Jeans length of $0.13~{\rm~pc}$. 
This result is similar to the findings of \citet{Rathborne2015}, who calculated the Jeans length of the Brick to be $0.10$~pc, by assuming an average volume density of $\sim 10^5 \mathrm{cm}^{-3}$. \citet{Rathborne2015} deduced that the break in the power spectrum of the Brick is indeed related to the Jeans length.
The break that we see in the simulations is also consistent with a Jeans-like scale, however it could be caused by other factors, such as the sonic scale or numerical dissipation. Because the current paper focuses on structural properties and a full kinematic analysis is needed to distinguish these interpretations, we will examine this in-depth in a follow-up study that focuses on the kinematic properties of the simulations.
%However, in the simulations we find a sonic scale of $\sim 0.4\pm0.1~\mathrm{pc}$ (see Petkova et al. in prep.), which matches well the break in the top panel of Figure~\ref{fig:power-spectrum}.

Finally, Table~\ref{tab:power-spec} shows that the synthetic emission maps have power spectrum slopes between $-4.2$ and $-2.3$ (if we exclude N$_2$H$^+$ with its high levels of noise). The group of molecules with compact emission (i.e.\ CS, HCN, HCO+ and HCN) have shallower slopes (above -3.5), while those that have more extended emission have steeper slopes (below -3.5). These slopes are broadly consistent with the results of \citet{Rathborne2015}, who find values about $-3$ for HCN, HCO$^+$ and HNCO (see also \citet{Henshaw2020}). Since the emission maps of \citet{Rathborne2015} consist of ALMA observations combined with single dish data, their power spectra were measured over a greater range of spatial scales than ours, and hence their slopes are more robust. This difference in available power law fitting range can account for some of the observed slope discrepancies.

\subsection{Fractal dimension}
\label{sec:fractal-dim}
Gas clouds exhibit self-similarity on a range of size scales. However, the self-similar density profile of a star-forming core appears very different from the self-similar structure of a supernova remnant. The former has smooth, spherical density contours, while the latter has a complex, turbulent structure. This difference can be captured by the concept of fractal dimension, which corresponds to the level of complexity within a structure.

We have calculated the fractal dimension of the synthetic emission maps of the Brick, using two different methods --- the perimeter-area method and the box-counting method. Both of these methods consider contours at different pixel thresholds within an emission map, and extract the fractal dimension from their properties. Note that here we use the 2D fractal dimension, since we measure it using 2D emission maps. Therefore, the fractal dimension is found within the range between 1 and 2, where 1 corresponds to smooth, circular contours, and 2 to high structural complexity.

\subsubsection{Perimeter-area method}
\label{sec:perim-area}

\begin{figure}
	\includegraphics[width=\columnwidth]{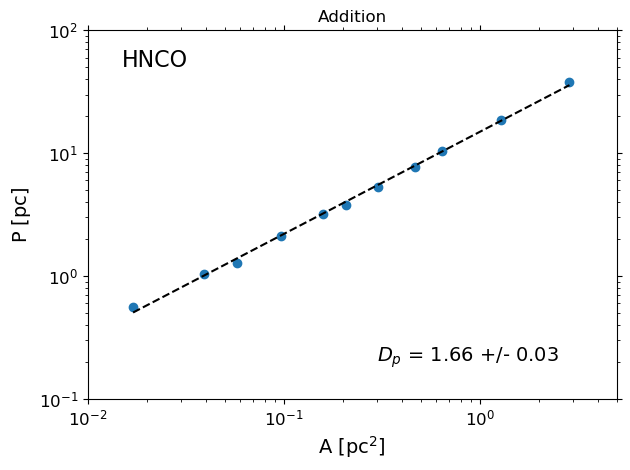}
	\includegraphics[width=\columnwidth]{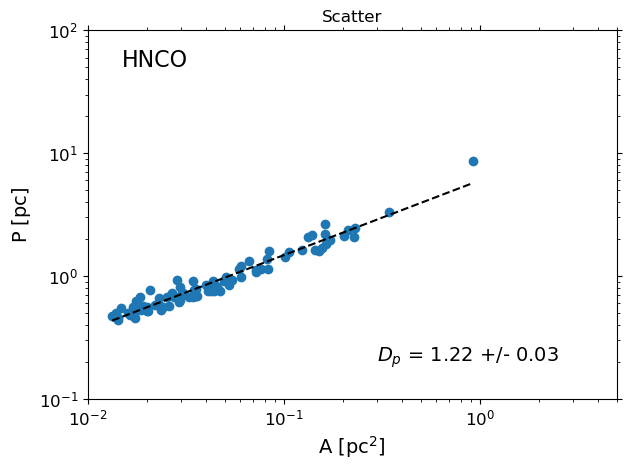}
	\caption{Fractal dimension, $D_p$, of the HNCO emission map of the TVir snapshot, obtained with the perimeter-area method, using the addition \textit{(top)} and the scatter \textit{(bottom)} interpretation.}
    \label{fig:perimeter-area-D}
\end{figure}

\begin{table*}
	\centering
	\caption{Fractal dimension, $D_p$, computed with the perimeter-area method for the simulated and observed emission maps of the Brick. Note that we have only included the observational results for those molecules from \citet{Rathborne2015}, which were also used with the simulation. The results reported by \citet{Rathborne2015} follow the addition interpretation.}
	\label{tab:fractal-dim}
	\begin{tabular}{lccccccr} % four columns, alignment for each
		\hline
		Molecular  & \multicolumn{2}{|c|}{$D_p$ (TVir)} & \multicolumn{2}{|c|}{$D_p$ (SVir)}& \multicolumn{3}{|c|}{$D_p$ (observed)}\\
		species & addition & scatter & addition & scatter & \cite{Rathborne2015} &addition & scatter\\
		\hline
		CS & $1.72\pm 0.04$ & $1.27 \pm 0.04$ & $1.80\pm 0.03$ & $1.36\pm 0.06$ &&&\\
		HCN & $1.84\pm 0.07$ & $1.15\pm 0.11$ & $1.74 \pm 0.05$ & $1.18 \pm 0.09$ & 1.5 & $1.50\pm 0.04$ & $1.28\pm 0.03$ \\
		HCO$^+$ & $1.70 \pm 0.04$ & $1.27 \pm 0.05$ & $1.80 \pm 0.03$ & $1.35 \pm 0.07$ & 1.4 & $1.29\pm 0.06$ & $1.25\pm 0.04$ \\
		HNCO& $1.66\pm 0.03$ & $1.22\pm 0.03$ & $1.67\pm 0.07$ & $1.23\pm 0.03$ & 1.5 & $1.48\pm 0.07$ & $1.25\pm 0.03$ \\
		H$_2$CS & $1.75\pm 0.06$ & $1.26\pm 0.04$ & $1.72\pm 0.04$ & $1.33\pm 0.03$ & $1.5$ & $1.91\pm 0.16$ & $1.31\pm 0.11$\\
		HC$_3$N & $1.64\pm 0.02$ & $1.26\pm 0.03$ & $1.71\pm 0.02$ & $1.34\pm 0.03$ &&&\\
		HNC & $1.84\pm 0.05$ & $1.11\pm 0.08$ & $1.73\pm 0.02$ & $1.34\pm 0.07$ &&&\\
		N$_2$H$^+$ & $1.81\pm 0.04$ & $1.30\pm 0.05$ & $1.87\pm 0.04$ & $1.44\pm 0.06$ &&&\\
		\hline
	\end{tabular}
\end{table*}

One way of calculating the fractal dimension of an observed cloud is by considering the area, $A$, and the perimeter, $P$, of different contours in the image. These two quantities are linked by the fractal dimension, $D_p$, in the following way \citep{Vogelaar1994}:

\begin{equation}
P \propto A^{D_p/2}.
\end{equation}
This means that an object with non-complex structure (such as a 2D-Gaussian) has $P \propto A^{1/2}$ and $D_p=1$, while an object filling up the space with the length of its winding contours has $D_p=2$.

In practice, we have cloud images consisting of pixels, which create limitations on our ability to accurately calculate $P$ and $A$. To obtain $A$ we select all pixels with values above a chosen level (that is varied) and we search for connected regions within that subset. A region is connected when each of its pixels shares a wall with at least one other pixel in the region. Counting the number of pixels in a connected region gives $A$, and $P$ is obtained by counting the number of pixels externally adjacent to the connected region (i.e.\ sharing a wall). Regions containing fewer than a chosen number of pixels are considered to be poorly resolved and hence we have not included them in the calculation of $D_p$. \citet{Vogelaar1994} demonstrate that this threshold should be at around 20 pixels. However, for an interferometric observation, the synthesised beam is the relevant resolution element, which is typically represented by multiple pixels. We limit the area of the smallest structures included in this measurement to be at least 4 times the area of the synthesised beam. This limit was empirically determined, based on selecting $(A,P)$ pairs that follow a single power law. In the case of the TVir and SVir emission maps, this size is 200 pixels, while in the real observations of the Brick, it is 65 pixels, due to differences in pixel sizes. Finally, the way in which we chose to determine $A$ and $P$ is not unique. Connected regions can include pixels sharing a diagonal border and not only a wall. The perimeter can include the sum of two pixel walls instead of just counting the pixel itself in parts where the boundary is diagonal. However, as demonstrated by \citet{Vogelaar1994} and \citet{Federrath2009}, tweaking the parameters of this method does not significantly change the value of $D_p$, but only the proportionality constant in the power law. 

We encounter two types of interpretations of the perimeter-area method found in the literature. One calculates the area and perimeter of each individual connected region and uses it as a data point to fit a line through (we refer to it as \textit{scatter} approach), while the other sums up the areas and perimeters of all connected regions for a given pixel level (we refer to it as \textit{addition} approach). Examples of the first interpretation can be seen in \citet{Vogelaar1994} and \citet{Sanchez2005}, and the second one was used by \citet{Rathborne2015}\footnote{\label{foot:fractal}Most authors do not explicitly state their interpretation of the perimeter-area method. Since the addition and scatter interpretations typically have visually distinct data sets (the former consist of monotonically increasing $(A,P)$ pairs, while the latter appears as scattered data points about a line; see Figure~\ref{fig:perimeter-area-D}), we use that as a diagnostic tool.}. More importantly, the two approaches are not equivalent and hence the fractal dimensions that they yield should not be compared to each other directly. This point can be illustrated if we consider a scenario of the scatter approach where all points lie on the line 
\begin{equation}
    \log(P) = \frac{D_p}{2} \log(A).
\end{equation}
By adding the areas and perimeters of two of the data points, we get a new data point which does not, in general, lie on the same line:
\begin{equation}
    \log(P_1+P_2) = \log(A_1^{D_p/2}+A_2^{D_p/2}) \neq \frac{D_p}{2} \log(A_1+A_2).
\end{equation}
On a more fundamental level, the difference between the interpretations resembles the distinction between \textit{intensive} and \textit{extensive} properties of matter. An intensive property is one which is scale free (such as temperature or pressure), while an extensive one depends on the length scale over which it is measured (such as mass or volume). The scatter approach yields an intensive property of the cloud, as its output would remain the same if we were to include an additional cloud in the emission map, with the same intrinsic fractal dimension as the original data. On the other hand, the addition approach can, under certain circumstances, produce a different result by including an additional structure (see Appendix ~\ref{sec:app-fractal-dim}). What separates the outcome of the addition method from a true extensive property, is that it is not additive, i.e.\ we cannot compute the fractal dimension in two sub-regions of the cloud and add them to obtain the total fractal dimension. Due to this fundamental discrepancy between the two interpretations, we recommend using the scatter approach when computing and comparing the fractal dimension of gas clouds.

In order to compare our emission maps to the existing literature, we compute the fractal dimension using both interpretations of the perimeter-area method. Figure~\ref{fig:perimeter-area-D} shows the addition and scatter interpretations, applied to the HNCO emission map of the TVir snapshot. We see that the addition interpretation produces a much steeper slope and hence a higher fractal dimension than the scatter interpretation. This behaviour continues for the remaining synthetic and real emission maps of the Brick (see summary of the results in Table~\ref{tab:fractal-dim}). We will consider the implications of the measured fractal dimensions further below.

\subsubsection{Box-counting method}
\begin{figure}
	\includegraphics[width=\columnwidth]{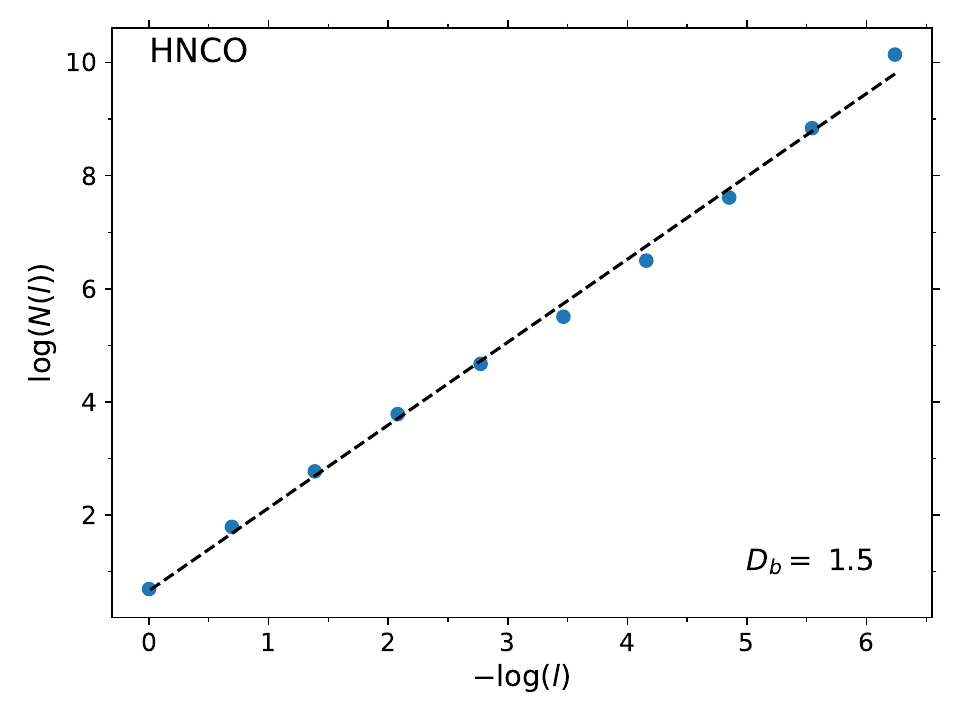}
	\includegraphics[width=\columnwidth]{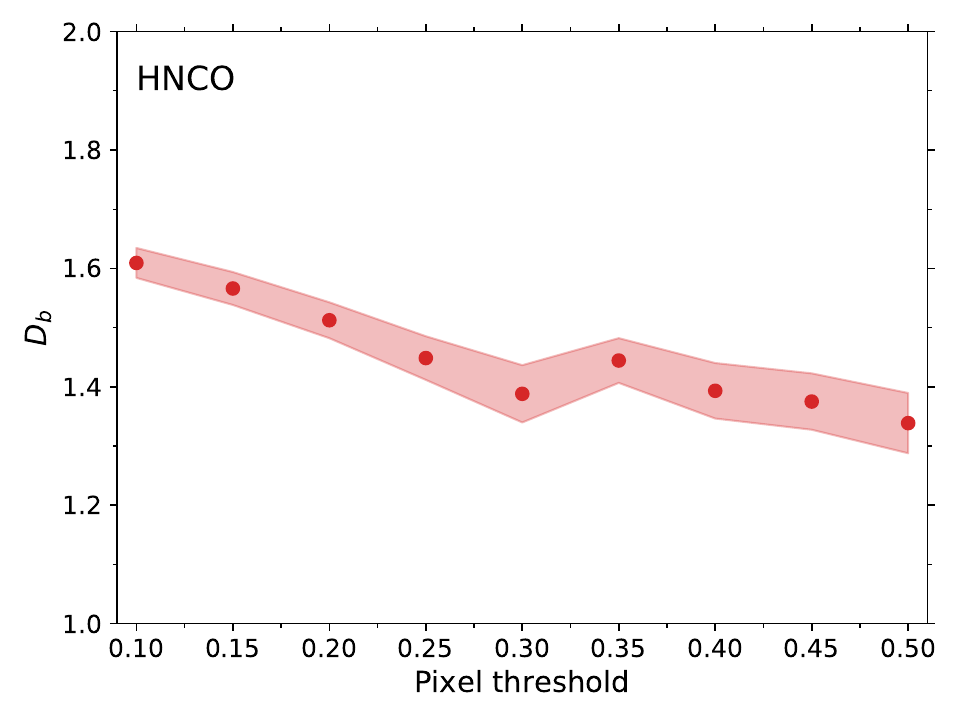}
    \caption{Fractal dimension, $D_b$, obtained with the box-counting method (see eq. ~\ref{eq:box-count}) for the HNCO emission map of the TVir snapshot. \textit{Top:} $D_b$ calculated at a pixel threshold given by the mean pixel value of the emission map ($\approx 15\%$ of the maximum pixel value). \textit{Bottom:} $D_b$ calculated for a range of pixel thresholds, which are given as a fraction of the maximum pixel count in the image. The shaded area gives a $1\sigma$ error bar to the slope of the linear fit for $D_b$.}
    \label{fig:box-counting}
\end{figure}

An alternative way of calculating the fractal dimension can be achieved by using the box-counting method. Similarly to the perimeter-area method, we select the pixels with values above a chosen threshold (specified below). We then cover the image with a grid of boxes and count the number of boxes, $N(l)$, containing the selected pixels as a function of box size, $l$. By varying $l$, we measure different values of $N(l)$, and for some range of $l$, $\log{N(l)}$ increases roughly linearly with $-\log{l}$. We can use that property to obtain the fractal dimension $D_b$, given by \citep{Federrath2009}:

\begin{equation}
N(l) \propto l^{-D_b}.
\label{eq:box-count}
\end{equation}

Unlike $D_p$, the fractal dimension obtained with the box-counting method depends on the pixel threshold. Figure~\ref{fig:box-counting} shows the $D_b$ fit for the HNCO emission map of the TVir snapshot. In the top panel of the figure we see $D_b$ fitted for a chosen pixel threshold, while in the bottom panel we see the change of $D_b$ as a function of pixel threshold. In order to compare $D_b$ between the different emission maps, we have fixed the threshold to the mean pixel value in the image (excluding the background pixels). Typically this threshold is around $\sim 10-20 \%$ of the maximum pixel value. The results of these calculations for the emission maps of all molecular species are shown in Table~\ref{tab:fractal-dim-box}.

\begin{table}
	\centering
	\caption{Fractal dimension of the TVir and SVir emission maps computed with the box-counting method. The pixel threshold is chosen to be the mean pixel value.}
	\label{tab:fractal-dim-box}
	\begin{tabular}{lcc} % four columns, alignment for each
		\hline
		Molecular species & $D_b$ (TVir) & $D_b$ (SVir)\\
		\hline
		CS & $1.48\pm 0.04$ & $1.40\pm 0.05$\\
		HCN & $1.36\pm 0.05$ & $1.26\pm 0.06$\\
		HCO$^+$ & $1.44\pm 0.05$ & $1.36\pm 0.05$\\
		HNCO& $1.53\pm 0.04$ & $1.41\pm 0.05$\\
		H$_2$CS & $1.45\pm 0.04$ & $1.37\pm 0.05$\\
		HC$_3$N & $1.54\pm 0.04$ & $1.45\pm 0.05$\\
		HNC & $1.43\pm 0.04$ & $1.30\pm 0.06$\\
		N$_2$H$^+$ & $1.25\pm 0.05$ & $1.10\pm 0.05$\\
		\hline
	\end{tabular}
\end{table}

\subsubsection{Fractal structure}
We find three different ranges of fractal dimension values for our synthetic emission maps, depending on the method used. The two different interpretations of the perimeter-area method yield $D_p = 1.6-1.9$ for the addition approach and $D_p = 1.2-1.4$ for the scatter approach. There is one outlier of the scatter approach, which is the HNC emission map of the TVir snapshot, with $D_p = 1.11$. Upon visual inspection of the image, we see that it consists of individual peaks, with close to circular contours, which explains the results. It is interesting to note that this image yields a very high fractal dimension of $1.84$ with the addition interpretation, which is an example of why we do not recommend using this approach. Additionally, we find the fractal dimension range from the box-counting method to be $D_b = 1.3-1.6$ (excluding the emission maps of N$_2$H$^+$, which are noisy and hence the level at which $D_b$ is calculated is closer to the peak brightness). There is no significant variation in fractal dimension between the TVir and SVir emission maps. In comparison, \citet{Rathborne2015} report a slightly lower fractal dimension $D_p = 1.4-1.7$ using the addition interpretation, with values between 1.4 and 1.5 for the four molecular tracers that overlap with our data set. However, we find inconsistent fractal dimension values for HCO$^+$ and H$_2$CS when we perform the measurement using the addition interpretation. This discrepancy is due to the fact that the addition interpretation is sensitive to the selection of structures within the image, and we exclude structures with areas that are smaller than four beam sizes. When using the scatter interpretation on the \citet{Rathborne2015} data, we find values within the range of $D_p = 1.2-1.3$, which are consistent with the synthetic maps.

The two distinct ranges of $D_p$ tell different stories about the fractal nature of the Brick. While the addition interpretation sees structures with relatively high level of complexity, the scatter interpretation sees a smoother cloud. From a methods perspective, we trust the results of the scatter interpretation, as discussed in Appendix ~\ref{sec:app-fractal-dim}. The values for interstellar gas clouds, obtained by other authors using this method, fall in the range of $D_p = 1.2-1.6$ \citep{Beech1987,Bazell1988,Dickman1990,Falgarone1991,Vogelaar1991,Hetem1993,Vogelaar1994,Sanchez2005}. This places the fractal dimension of both the observed and simulated emission maps of the Brick in the lower half of that range. 

The fractal dimension, given by the box-counting method, is harder to compare to literature values, since it is a function of the pixel threshold. Figure~\ref{fig:box-counting} shows that as we increase the pixel threshold, we gradually lower $D_b$ from about 1.6, down to about 1.4. This trend arises from the fact that by selecting only the brightest peaks, we consider small spatial scales, where we lack the sufficient resolution to see fractal substructure. Therefore, the highest value of $D_b$ serves as an upper estimate of the fractal nature of the synthetic emission maps.

All of the above results indicate that neither the observed, nor the simulated Brick cloud have unusual fractal structure, compared to typical interstellar gas clouds.

\subsection{Moments of inertia characterised with the J-plots method}
\label{sec:j-plots}
\begin{figure}
	\includegraphics[width=\columnwidth]{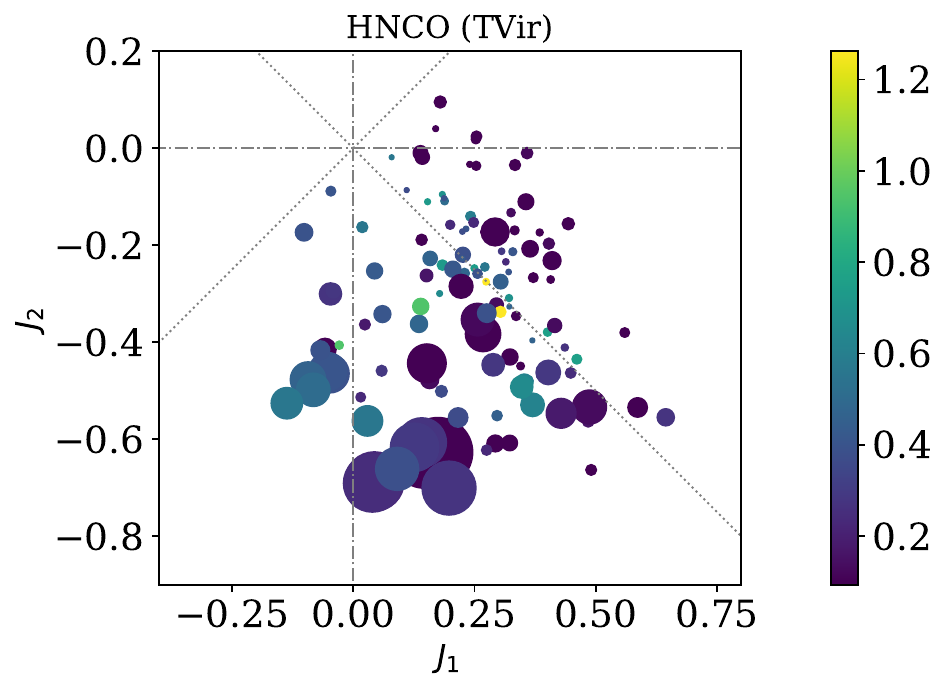}
	\includegraphics[width=\columnwidth]{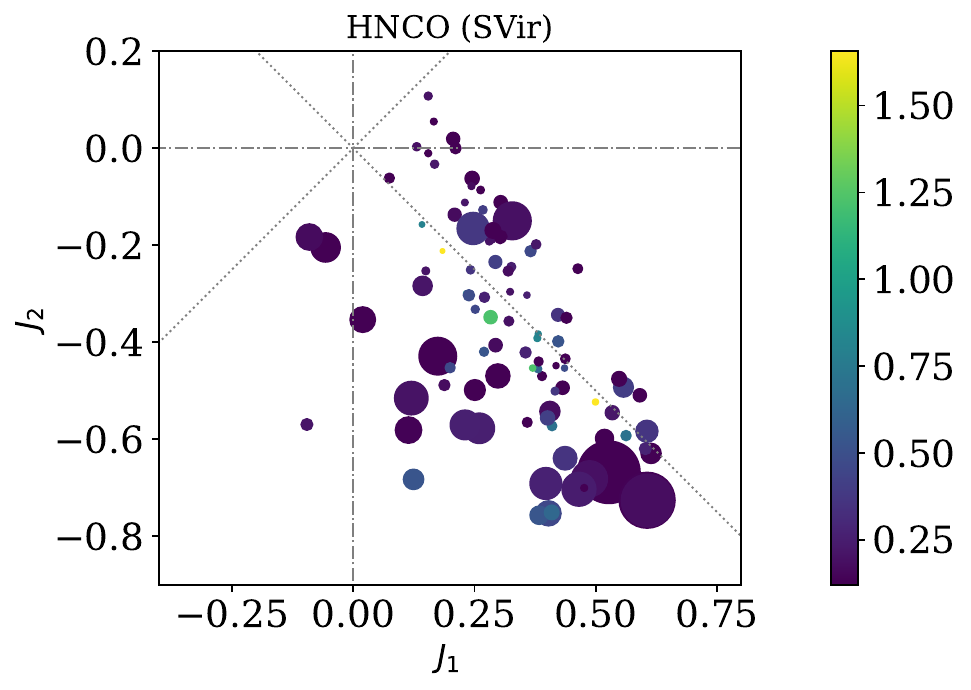}
	\includegraphics[width=\columnwidth]{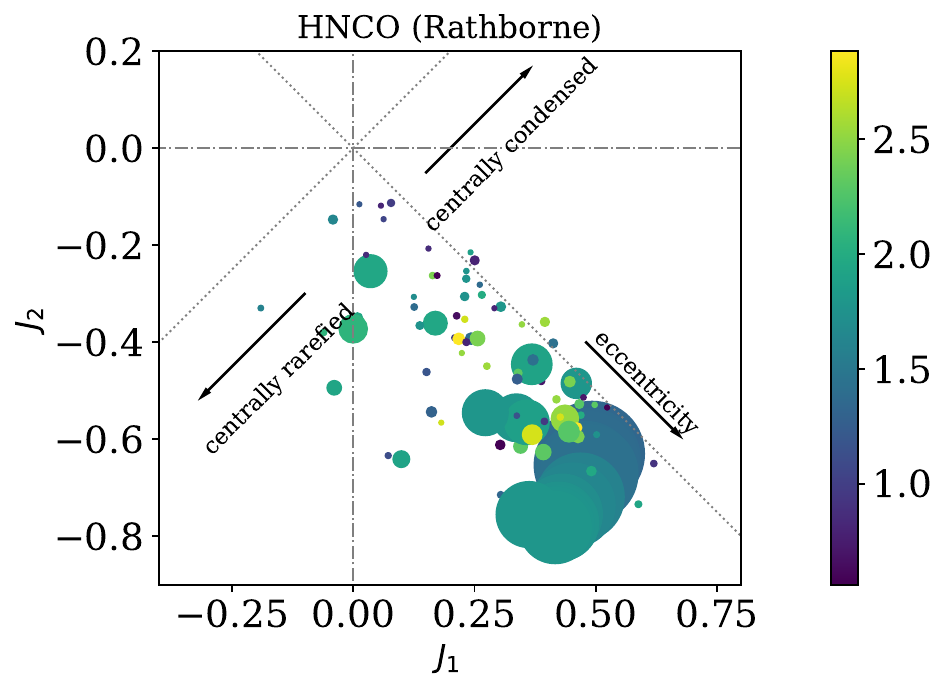}
	\caption{J plots of the HNCO emission maps, generated with \textsc{casa}, of the TVir (\textit{top}) and the SVir snapshot (\textit{middle}), together with the HNCO observation of \citet{Rathborne2015} (\textit{bottom}). Each datapoint corresponds to an identified structure within the corresponding dendrogram, and its size is proportional to the area of the structure. The colour indicates the intensity level at which each structure is identified in Jy/beam~km/s.}
    \label{fig:j-plot}
\end{figure}

We now quantify the cloud substructure by using the principal moments of inertia, characterised through the J-plots method \citep{Jaffa2018}. With this method, we can distinguish between centrally-condensed (e.g.\ cores), centrally-rarefied (e.g.\ rings), and elongated structures (e.g.\ filaments).

In order to apply the J-plots method to the emission maps generated with \textsc{casa}, we first identify 2D substructures in the maps by constructing dendrograms with the \textsc{astrodendro} Python package\footnote{http://www.dendrograms.org}. Each dendrogram structure is a contiguous group of pixels with values above a given threshold (similarly to the structures used for measuring the fractal dimension). The dendrogram is constructed starting from the brightest pixel and gradually decreasing the threshold to identify more structures and their hierarchical connections. The dendrogram construction is controlled by three parameters, which are the minimum pixel value (lowest possible threshold), the minimum delta (minimum threshold distance between two hierarchically connected structures), and the minimum number of pixels in a structure. For our analysis, we set the minimum value to three times the root-mean-squared (RMS) noise of the image background. Note that the RMS is computed for an empty region at the periphery of each image. The minimum number of pixels must exceed the beam size, as the latter defines our resolution. For the synthetic observations we have used 200 as the minimum number of pixels (with the beam area covering $\approx 25$ pixels), and for the real observations we have used 65 pixels. This accounts for differences in the pixel size between the two sets of images, and it ensures that the minimum number of pixels cover approximately the same physical area. The minimum delta is set to 10\% of the brightest peak.

After constructing the dendrogram, we compute the J moments of each structure in the following way, as described by \citet{Jaffa2018}. Let a structure consist of $P$ pixels, with values $\Sigma_p$, where $p\in[1,P]$. The J plot method treats $\Sigma_p$ as surface density, and hence uses the terms ``mass'' and ``centre of mass''. We will adhere to this language for consistency with \citet{Jaffa2018}, but note that in our calculations $\Sigma_p$ is the intensity.

The area, $A$, and mass, $M$, of the structure described above are given respectively as:
\begin{equation}
    A\ = P\ \Delta A, \ \ M\ = \sum_{p=1}^{P} \{\Sigma_{p}\}\ \Delta A,
\end{equation}
where $\Delta A$ is the area of a pixel. We can then compute the moments $M_{0}=M,\ M_{x},\ M_{y},\ M_{xx},\ M_{yy},\ M_{xy}$, using the position of each pixel center, $(x_{p}, y_{p})$,\footnote{E.g.\ $M_{x}= \Delta A \sum_{p=1}^{P}{\Sigma_p x_p}$ and $M_{xy}= \Delta A \sum_{p=1}^{P}{\Sigma_p x_p y_p}$.} which leads to the center of mass:
\begin{equation}
    X\ =\ \frac{M_{x}}{M_{0}}, \ \ \ Y\ = \ \frac{M_{y}}{M_{0}}.
\end{equation}
The moments about the Cartesian axes are
\begin{equation}
    I_{xx}=M_{xx}-M_{x}X, \ \ \ I_{yy}=M_{yy}-M_{y}X \ \ \mathrm{and}\  I_{xy}=M_{xy}-M_{0}XY.
\end{equation}
The principal moments are then given by
\begin{equation}
    I_{1,2}=\left(\frac{I_{xx}+I_{yy}}{2}\right) \mp \left\{\left(\frac{I_{xx}+I_{yy}}{2}\right)^{2}-\left(I_{xx}I_{yy}-I_{xy}^{2}\right) \right\}^{\frac{1}{2}},
\end{equation}
where the indices $\{1,2\}$ refer to the different signs ($\mp$). From this definition, the J moments are defined for $i\ =\ 1,2$:
\begin{equation}
    J_{i}=\ \frac{I_{0}-I_{i}}{I_{0}+I_{i}},
\end{equation}
where $I_{0}=A M/4 \pi$. Note that we always have $J_1 \geq J_2$.

After computing the J moments, we plot $J_2$ against $J_1$ and we interpret the J plot in the following way. The position of a datapoint projected along the $J_1=J_2$ diagonal determines the degree of central condensation/rarefaction of the structure, with negative values of $J_1$ and $J_2$ corresponding to a centrally-rarefied, and positive values of $J_1$ and $J_2$ corresponding to a centrally-condensed structure. Additionally, the projected position along the $J_2=-J_1$ diagonal indicates the aspect ratio of the structure, with $J_2=J_1=0$ being a circle, and $J_1>0$, $J_2<0$ being an elongated structure. In this context, a curved filament would count as a (somewhat) centrally-rarefied structure.

Figure~\ref{fig:j-plot} shows the J plots of the synthetic (TVir: \textit{top panel}; SVir: \textit{middle panel}) and real (\citealt{Rathborne2015}: \textit{bottom panel}) integrated HNCO emission maps of the Brick. The size of each data point is proportional to the physical area of the corresponding structure, and the colour indicates the pixel threshold at which the structure has been identified within the dendrogram. The structures in Figure~\ref{fig:j-plot} occupy the area of the J plot around the $J_2=-J_1$ diagonal where $J_1\ga0$ and $J_2\la0$. The TVir and SVir structures are approximately evenly distributed on both sides of the diagonal, indicating a balance between centrally condensed and rarified structures. Interestingly, the centrally-condensed structures ($J_2>-J_1$) are predominantly close to circular, whereas the centrally-rarefied structures ($J_2<-J_1$) are predominantly elongated. By contrast, the structures in the \citet{Rathborne2015} observations are almost entirely below the diagonal, indicating that the real Brick is lacking centrally-condensed structures. None of the three data sets exhibit a preferred elongation of the structures, because the orthogonal projections of the data points on the $J_2=-J_1$ diagonal are approximately uniformly distributed across the spanned range of values. Finally, there is no apparent correlation between the size or the colour of a data point and its position on the J plot of the simulated emission maps. In the J plot of the real Brick we see that the largest data points correspond to structures that are elongated and slightly centrally-rarefied. This is due to the fact that on a cloud scale, the Brick is distinctly bean-shaped.

The above described difference in the centrally-condensed regions of the simulated and the real Brick also persists for the other molecular tracers.
This difference may be understood in terms of the ongoing process of star formation. The simulations are actively undergoing fragmentation and gravitational collapse, and by the time of the selected snapshots, close to half of the original gas mass has been transformed into sink particles, which are fed from condensations of locally gravitationally-bound gas. These gravity-dominated structures appear as the centrally-condensed data points in the J plot. The reason why we do not see centrally-condensed structures in the J plot of the real Brick is likely because its gravity-dominated regions are less numerous (since the Brick appears primarily quiescent) and smaller than the simulated ones. Indeed, \citet{Walker2021} have detected centrally-condensed star-forming cores in the Brick, with sizes of $\sim 1000~{\rm AU}$, but only few compared to the simulations. In order for these cores to be visible on the J plot, we require a much higher resolution than that of the observations of \citet{Rathborne2015}.

The J plot results of Figure~\ref{fig:j-plot} reflect differences in the star formation rate and star formation size scale between the simulations and the observations. We have already discussed the differences in length scales relevant for star formation in Section~\ref{sec:power-spectrum}, so we restrict the brief discussion here to the star formation rate. \citet{Federrath2016} predicted the star formation rate per free-fall time of the Brick to be $\epsilon_{\rm ff} = 0.042 \pm 0.030$, using the star formation model of \citet{Krumholz2005} with the parameter fits from \citet{Federrath2012}. A later measurement by \citet{Barnes2017} found $\epsilon_{\rm ff} = 0.02$, which is in agreement with the theoretical prediction. 
We use the same theoretical models as \citet{Barnes2017} to predict the the star formation rate per free-fall time of the TVir simulation for three different star formation theories. By using our values for $\mathcal{M}$, $b$ and $\beta$ (see Section~\ref{sec:pdfs}), and the virial parameter of the simulated cloud at the analysed snapshot ($\alpha_{\rm vir} = 1.24$, \citealt{Dale2019}), we obtain $\epsilon_{\rm ff, TVir} = \{0.0021, 0.034, 0.298\}$ for the models of \citet{Krumholz2005}, \citet{Padoan2011} and \citet{Hennebelle2013}, respectively. This gives us a large spread of possible values due to differences in the star formation theories. In reality, the TVir simulation is actively forming stars, with $\epsilon_{\rm ff}=0.39$ \citep{Dale2019}, and we need to slow this process down in order to match the observations. One way to achieve this is by including magnetic fields. Indeed, if we assume the same turbulent magnetic field as is present in the real Brick (with $\beta=0.34$), the predicted values for $\epsilon_{\rm ff, TVir}$ for the three star formation theories become considerably smaller, with $\epsilon_{\rm ff, TVir} = \{0.0001, 0.023, 0.004\}$.
%This is still significantly larger than the predicted and calculated values of $\epsilon_{\rm ff}$ for the Brick (although the large uncertainties may account for some of this difference). The remainder of the difference reflects small differences in the Mach number and virial parameter. 
This leads us to the prediction that a magneto-hydrodynamic simulation of our Brick model should better reproduce both the observed star formation rate and the multi-scale structure of the real Brick quantified in Figure~\ref{fig:j-plot}.

\section{Conclusions}
\label{sec:conclusion}
We have presented a detailed study of the complex, multi-scale structure of numerical simulations and ALMA observations of the CMZ cloud G0.253+0.016 (`the Brick'). To do so, we have created synthetic emission maps of the Brick, using snapshots of two hydrodynamics simulations \citep{Dale2019,Kruijssen2019}, which have been post-processed with the line radiative transfer code \textsc{polaris} and subsequently with \textsc{casa}. The chosen snapshots match the current position of the Brick along its orbit. We have adopted the observational setup of \citet{Rathborne2015} for our post-processing to enable a direct comparison between their observations of the real Brick and our numerically modelled ones.

We have produced emission maps of eight molecular species per simulation snapshot (TVir and SVir), tracing gas of different densities. We find that half of our molecules (CS, HCN, HCO$^+$ and HNC) produced emission maps with compact structures, while the other half (HNCO, H$_2$CS, HC$_3$N and N$_2$H$^+$) traced more diffuse gas. Furthermore, all of our molecular emission maps show at least moderate correlation with the gas surface density. This is in contrast to the real Brick, where \citet{Rathborne2015} found poor correlation between their molecular emission maps and the dust continuum, which is assumed to trace the column density. This discrepancy between the simulations and the observations is likely due to the simplified assumptions of constant temperature and abundances in our modelling, as well as due to potential structural differences between the simulations and the real Brick.

We have studied the structure of our simulated emission maps by constructing brightness and density PDFs and power spectra, and by calculating the fractal dimensions of substructures and their moments of inertia through the J-plots method. The density PDFs of the simulations have a larger width than those of the real Brick, which is likely explained by the lack of magnetic fields in the simulations. Indeed, assuming the observed value of the turbulent plasma beta ($\beta=0.34$) in our simulations compensates for the difference in density dispersion.

We have constructed the spatial power spectra of the column density and the synthetic emission maps to study the turbulent structure of the cloud. We find that the column density power spectrum has a break at $\sim 0.32$~pc, which is
consistent with the plausible range for the Jeans length in the simulations ($0.1{-}0.5$~pc).
%consistent with the sonic scale. 
The large length scale slope of the column density power spectrum is similar to, but slightly lower than the value obtained by \citet{Rathborne2015}. The slightly shallower slope may be caused by the lack of magnetic fields in the hydrodynamics simulation. The power spectra of the synthetic emission maps follow a single power law within their reliable fitting range of spatial scales. We find slopes in the range between $-4.2$ and $-2.3$, which is approximately consistent with the findings in \citet{Rathborne2015}, who reported slopes of $-3$.

We have studied the fractal structure of the synthetic emission maps, using the perimeter-area method and the box-counting method. We find good agreement between the synthetic ($D_p = 1.2-1.4$) and the observed emission maps ($D_p = 1.2-1.3$), when using the perimeter-area method. We obtain $D_b \approx 1.3-1.6$ with the box-counting method, applied to the synthetic emission maps. These results indicate that the Brick and its simulated counterpart have typical fractal structure for an interstellar gas cloud. 

Additionally, we report that we have identified two interpretations of the perimeter-area method in the literature. In one of them individual connected regions within a contour are treated as separate entities, while in the other their areas and perimeters are added up. We advise against the interpretation involving addition, as discussed in Appendix~\ref{sec:app-fractal-dim}.

Finally, we have studied the moments of inertia of individual cloud substructures, by applying the J-plots method \citep{Jaffa2018}. Through this method we can distinguish between centrally-condensed (cores), centrally-rarefied (rings) and elongated structures (filaments). We find that the synthetic emission maps contain a mixture of centrally-condensed and centrally-rarefied structures with varying degrees of elongation. In contrast, the real Brick consists of almost exclusively centrally-rarefied structures. We attribute these differences to the ongoing process of star formation and in differences in the star formation rate between the simulations and the real Brick.

Throughout our analysis, we have seen consistent similarities between the TVir and SVir snapshots, even though \citet{Dale2019} find that the tidally-virialised simulations are better models of the CMZ clouds than the self-virialised ones, based on the more rapid gravitational collapse of the latter. This indicates that the small scale properties of the gas are not strongly influenced by the global amount of initial kinematic support of the cloud (at least at the orbital position of the Brick), and it makes our results more robust against variations in the initial conditions of the clouds.

Our results demonstrate that the underlying Brick simulations reproduce some important structural properties of the cloud \citep[in addition to the integrated properties studied by][]{Kruijssen2019}, such as the hierarchical substructure and the fractal dimension. This implies that many aspects of the cloud's evolution are affected by its orbit through the gravitational potential in the Galactic Centre region. In order to reproduce other observables, such as the power spectrum and the weakly correlated emission maps, we require
more complex models. These may include a more thorough treatment of the chemistry, and the addition of magnetic fields in the simulations. Models that include the external gravitational potential of the Galactic Centre take the first necessary step towards our understanding of the complex interplay between turbulence and self-gravity within this region.
%the addition of magnetic fields and a more careful treatment of the chemistry. Models that include the external gravitational potential generated by the stars in the CMZ better reproduce the observed structure, highlighting that cloud structure in the CMZ results from the complex interplay between internal physics (turbulence, self-gravity, magnetic fields) and the impact of the extreme environment in the Galactic centre region.

\section*{Acknowledgements}
We thank an anonymous referee for a timely and constructive report.

MAP would like to thank her colleagues from the MUSTANG group for many discussions and assistance throughout this project.

MAP and JMDK gratefully acknowledge funding from the European Research Council (ERC) under the European Union's Horizon 2020 research and innovation programme via the ERC Starting Grant MUSTANG (grant agreement number 714907).
MAP, JMDK, SCOG, and SR acknowledge financial support from the Deutsche Forschungsgemeinschaft (DFG; German Research Foundation) via the collaborative research center (SFB 881, Project-ID 138713538) ``The Milky Way System'' (MAP, JMDK, SCOG, SR: subproject B2; SCOG, SR: subprojects B1 and B8; SCOG: subproject A1).
MAP acknowledges support from a Chalmers Cosmic Origins postdoctoral fellowship.
JMDK gratefully acknowledges funding from the DFG through an Emmy Noether Research Group (grant number KR4801/1-1).
COOL Research DAO is a Decentralised Autonomous Organisation supporting research in astrophysics aimed at uncovering our cosmic origins.
SCOG acknowledges subsidies from the Heidelberg Cluster of Excellence STRUCTURES in the framework of Germany's Excellence Strategy (grant EXC-2181/1 - 390900948) and funding from the ERC via the ERC Synergy Grant ECOGAL (grant 855130).
DLW gratefully acknowledges support from the National Science Foundation under Award No. 1816715. 
SR acknowledges support from the DFG in the Priority Program SPP 1573 "Physics of the Interstellar Medium" (grant numbers KL 1358/18.1, KL 1358/19.2). SR also acknowledges access to computing infrastructure support by the state of Baden-W{\"u}rttemberg through bwHPC and the German Research Foundation (DFG) through grant INST 35/1134-1 FUGG.

This research made use of astrodendro (http://www.dendrograms.org), a Python package to compute dendrograms of Astronomical data, and Astropy,\footnote{http://www.astropy.org} a community-developed core Python package for Astronomy \citep{astropy:2013, astropy:2018}.

\section*{Data availability}

The data underlying this article will be shared on reasonable request to the corresponding author.

%%%%%%%%%%%%%%%%%%%%%%%%%%%%%%%%%%%%%%%%%%%%%%%%%%



\bibliographystyle{mnras}
\bibliography{references} % if your bibtex file is called references.bib

%%%%%%%%%%%%%%%%%%%%%%%%%%%%%%%%%%%%%%%%%%%%%%%%%%

%%%%%%%%%%%%%%%%% APPENDICES %%%%%%%%%%%%%%%%%%%%%

\appendix

\section{The integral of a cubic spline kernel}
\label{sec:app-density}
\begin{figure}
	\includegraphics[width=\columnwidth]{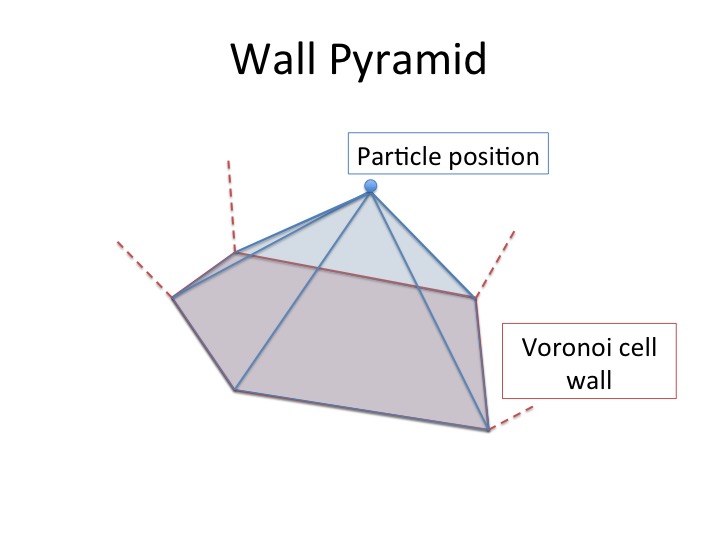}
	\includegraphics[width=\columnwidth]{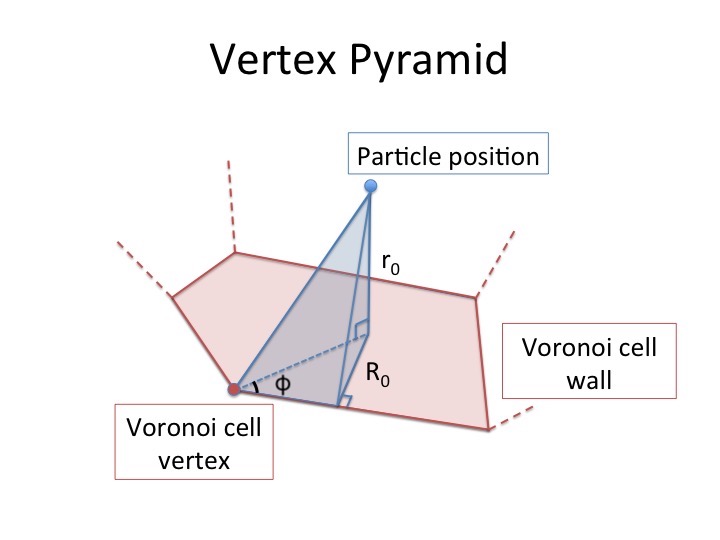}
    \caption{A wall pyramid and a vertex pyramid representations from \citet{Petkova2018}.}
    \label{fig:pyramids}
\end{figure}

In order to accommodate for the irregular shape of a Voronoi cell, the integration volume is divided into pyramids, as shown in Figure~\ref{fig:pyramids}. First the vertices of each cell wall are connected to the particle position, creating a wall pyramid, and then each wall pyramid is divided into vertex pyramids. Each vertex pyramid is characterised by the distance from the particle to the wall, $r_0$, the distance from the orthogonal projection of the particle position on the plane of the wall to an edge of the wall, $R_0$, and an angle $\phi$, as shown in Figure~\ref{fig:pyramids}.

The integral of $W$ over the volume of a vertex pyramid can be written as:

\begin{equation}
   I_{P}= \frac{r_0^3}{h^3\pi}  \begin{cases}
      \frac{1}{6} I_{-2} - \frac{3}{40} \left (\frac{r_0}{h} \right )^2 I_{-4} + \frac{1}{40} \left (\frac{r_0}{h} \right )^3 I_{-5} \\+ \frac{B_1}{r_0^3} I_0, & \frac{r_0}{h}\leq \mu; \\
       & \\
      \frac{1}{4} (\frac{4}{3} I_{-2} - (\frac{r_0}{h}) I_{-3}+ \frac{3}{10} (\frac{r_0}{h})^{2} I_{-4} \\- \frac{1}{30} (\frac{r_0}{h})^{3} I_{-5} + \frac{1}{15} (\frac{r_0}{h})^{-3} I_1) + \frac{B_2}{r_0^3} I_0, & \frac{r_0}{2h}\leq \mu\leq \frac{r_0}{h}; \\
       & \\
      - \frac{1}{4} (\frac{r_0}{h})^{-3} I_1 + \frac{B_3}{r_0^3} I_0, & \mu\leq \frac{r_0}{2h}, \\
  \end{cases} 
  \label{eq:IP3D}
\end{equation}
where

\begin{equation}
I_0 = \phi + C,
\end{equation}

\begin{equation}
I_{-2} = \phi +  \left (\frac{R_0} {r_0}\right )^2 \tan \phi + C,
\end{equation}

\begin{equation}
I_{-4} = \phi +  2\left (\frac{R_0} {r_0} \right )^2\tan \phi + \frac{1}{3}\left (\frac{R_0} {r_0}\right )^4 \tan \phi (\sec^2 \phi + 2) + C,
\end{equation}

\begin{equation}
I_{1} = \tan^{-1} \left ( \frac{u}{\alpha} \right) + C,
\end{equation}

\begin{multline}
I_{-3} = \frac{\alpha(1+\alpha^2)}{4} \left ( \frac{2u}{1-u^2} + \log(1+u) - \log(1-u) \right ) \\ 
+ \frac{\alpha}{2} ( \log(1+u) - \log(1-u) ) + \tan^{-1} \left ( \frac{u}{\alpha} \right) + C,
\end{multline}

\begin{multline}
I_{-5} = \frac{\alpha(1+\alpha^2)^2}{16} \left ( \frac{10u-6u^3}{(1-u^2)^2} + 3(\log(1+u) - \log(1-u)) \right ) \\ 
+ \frac{\alpha(1+\alpha^2)}{4} \left ( \frac{2u}{1-u^2} + \log(1+u) - \log(1-u) \right ) \\ 
+ \frac{\alpha}{2} ( \log(1+u) - \log(1-u) ) + \tan^{-1} \left ( \frac{u}{\alpha} \right) + C,
\end{multline}

\begin{equation}
B_1 = \frac{r_0^3}{4} \left ( -\frac{2}{3} + \frac{3}{10} \left ( \frac{r_0}{h} \right )^2 - \frac{1}{10} \left ( \frac{r_0}{h} \right )^3 \right );
\end{equation}

\begin{equation}
   B_2 = \frac{r_0^3}{4}  \begin{cases}
      -\frac{2}{3} + \frac{3}{10} \left ( \frac{r_0}{h} \right )^2 - \frac{1}{10} \left ( \frac{r_0}{h} \right )^3 - \frac{1}{5} \left ( \frac{r_0}{h} \right )^{-2}, & r_0\leq h; \\
       & \\
      -\frac{4}{3} + \left ( \frac{r_0}{h} \right ) - \frac{3}{10} \left ( \frac{r_0}{h} \right )^2 + \frac{1}{30} \left ( \frac{r_0}{h} \right )^3  \\- \frac{1}{15} \left ( \frac{r_0}{h} \right )^{-3}, & h\leq r_0\leq 2h; \\
  \end{cases}
\end{equation}

\begin{equation}
   B_3 = \frac{r_0^3}{4}  \begin{cases}
      -\frac{2}{3} + \frac{3}{10} \left ( \frac{r_0}{h} \right )^2 - \frac{1}{10} \left ( \frac{r_0}{h} \right )^3 + \frac{7}{5} \left ( \frac{r_0}{h} \right )^{-2}, & r_0\leq h;  \\
       & \\
      -\frac{4}{3} + \left ( \frac{r_0}{h} \right ) - \frac{3}{10} \left ( \frac{r_0}{h} \right )^2 + \frac{1}{30} \left ( \frac{r_0}{h} \right )^3  \\- \frac{1}{15} \left ( \frac{r_0}{h} \right )^{-3} + \frac{8}{5} \left ( \frac{r_0}{h} \right )^{-2}, & h\leq r_0\leq 2h; \\
       & \\
      \left ( \frac{r_0}{h} \right )^{-3},  &  r_0\geq 2h. \\
  \end{cases} 
\end{equation}
In the above $\alpha=\frac{R_0}{r_0}$, $u=\sqrt{1-(1+\alpha^2) \mu^2}$, and $\mu=\cos{\phi}$.

Even though the use of the analytic integral greatly speeds up the density and velocity mapping, there are still multiple expensive calculations that need to be performed. For this reason, we have pre-computed and tabulated the values of $I_P$ for various configurations of $r_0$, $R_0$ and $\phi$. Any desired value of $I_P$ is then obtained by linearly interpolating between the pre-computed values. The tabulated approach introduces a small error in the mapped SPH parameters, which has been has been shown to be within 0.6\% \citep{Petkova2018b}.

\section{Moment maps}
\label{sec:app-moments}

\begin{figure*}
	\includegraphics[width=\textwidth]{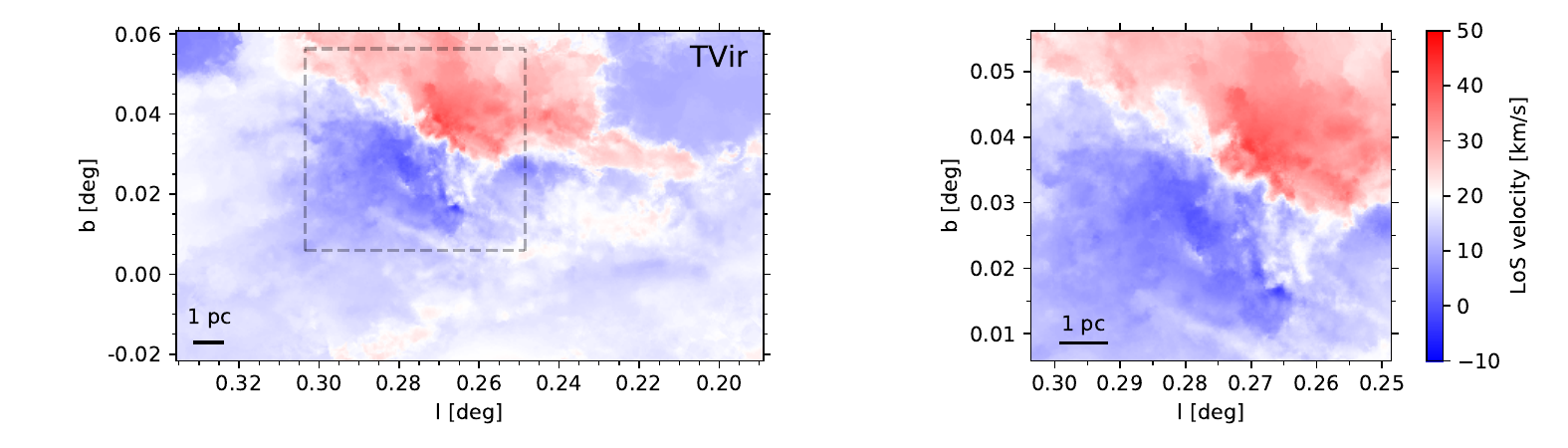}
	\includegraphics[width=\textwidth]{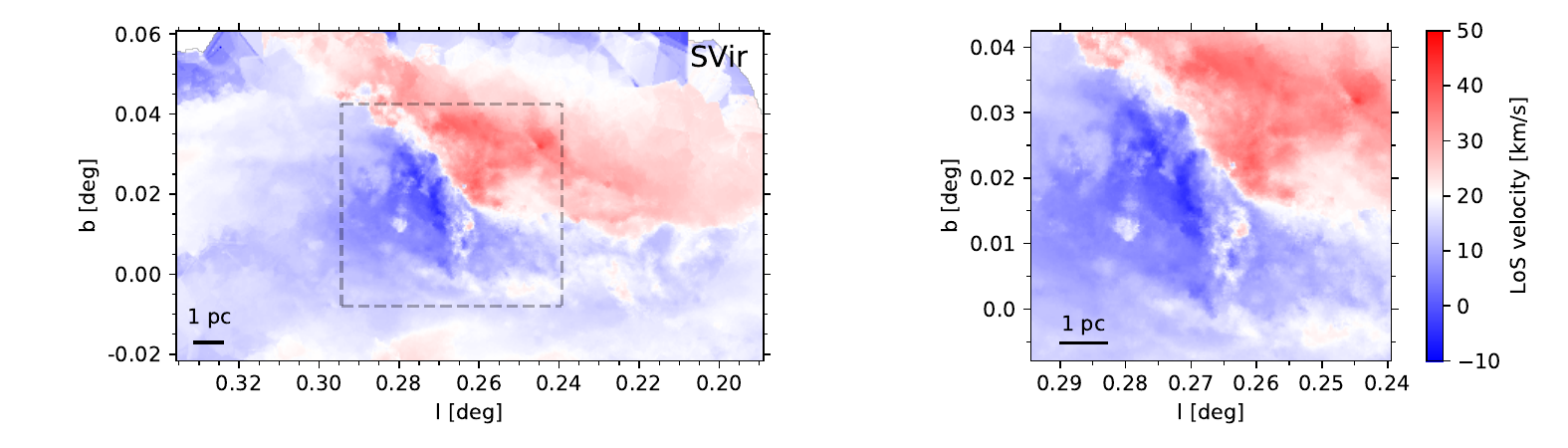}
	\caption{HNCO intensity-weighed velocity map of the full simulated clouds (\textit{left}), and the high-density regions, selected as the observation areas for the synthetic ALMA emission maps. (\textit{right}). The top panels show the TVir snapshot, and the bottom ones show the SVir snapshot. Note that there is a small offset between the positions of the right-hand panels (also shown as dashed rectangles on the left), which comes from the lopsided mass distribution in each cloud. The colour bars indicate the line-of-sight velocities in each of the panels.}
    \label{fig:mom1}
\end{figure*}

\begin{figure*}
	\includegraphics[width=\textwidth]{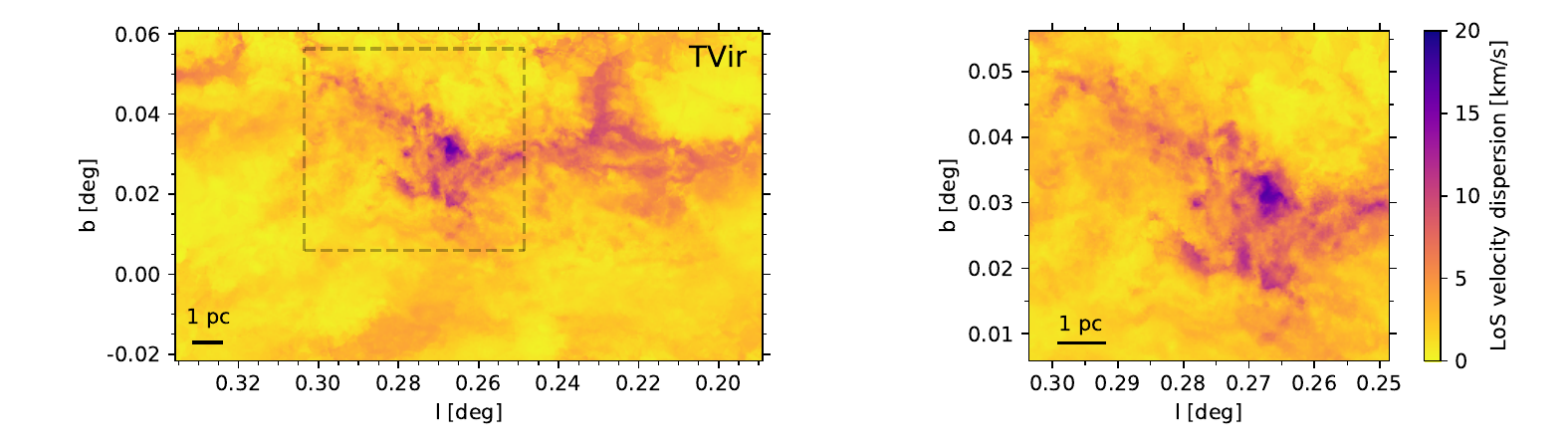}
	\includegraphics[width=\textwidth]{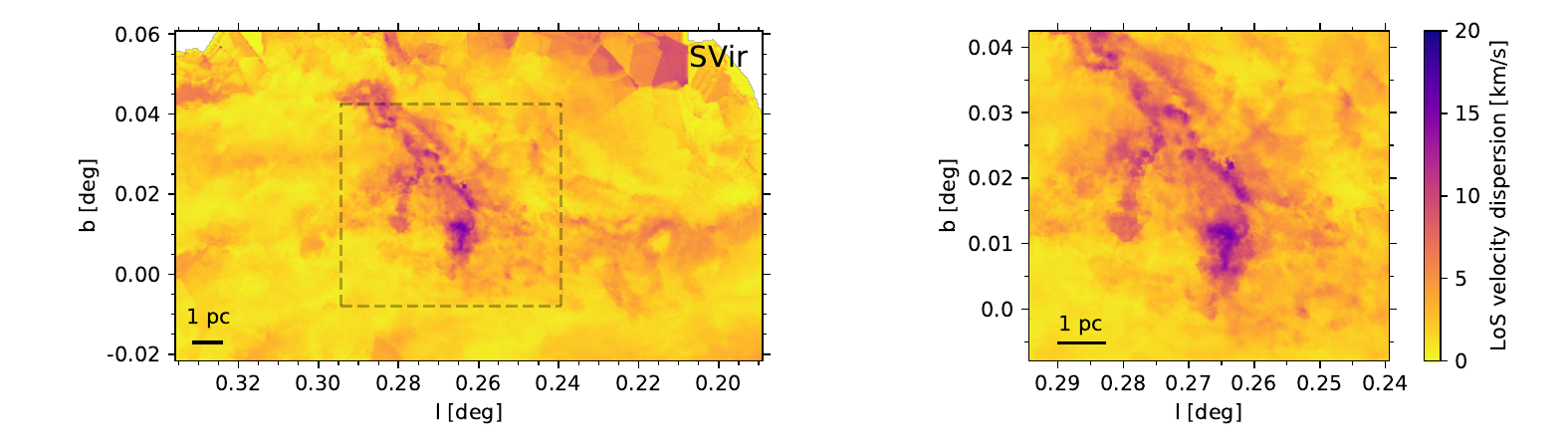}
	\caption{Second velocity moment map of HNCO for the full simulated clouds (\textit{left}), and the high-density regions, selected as the observation areas for the synthetic ALMA emission maps. (\textit{right}). The top panels show the TVir snapshot, and the bottom ones show the SVir snapshot. Note that there is a small offset between the positions of the right-hand panels (also shown as dashed rectangles on the left), which comes from the lopsided mass distribution in each cloud. The colour bars indicate the line-of-sight velocity dispersions in each of the panels.}
    \label{fig:mom2}
\end{figure*}

In addition to the other results, we include the first and second velocity moment maps of HNCO for the two simulation snapshots (see Figure \ref{fig:mom1} and \ref{fig:mom2} respectively). In order to construct these maps, we combine the velocity channel data that is directly produced by \textsc{polaris} (without further processing with \textsc{CASA}). Figure \ref{fig:mom1} shows that both snapshots present clear signs of rotation, with a very similar velocity range. These velocities are an excellent match to the real Brick, which has line-of-sight velocities between $\sim 0~\rm{km~s^{-1}}$ and $\sim 45~\rm{km~s^{-1}}$ \citep{Rathborne2015,Federrath2016,Henshaw2019}. The velocity dispersions in Figure \ref{fig:mom2} are also similar between the two snapshots, with values up to $\sim 20~\rm{km/s}$ for a pixel size of $0.04~\rm{pc}$.

\section{Interpretations of the perimeter-area method}
\label{sec:app-fractal-dim}

\begin{figure}
	\includegraphics[width=\columnwidth]{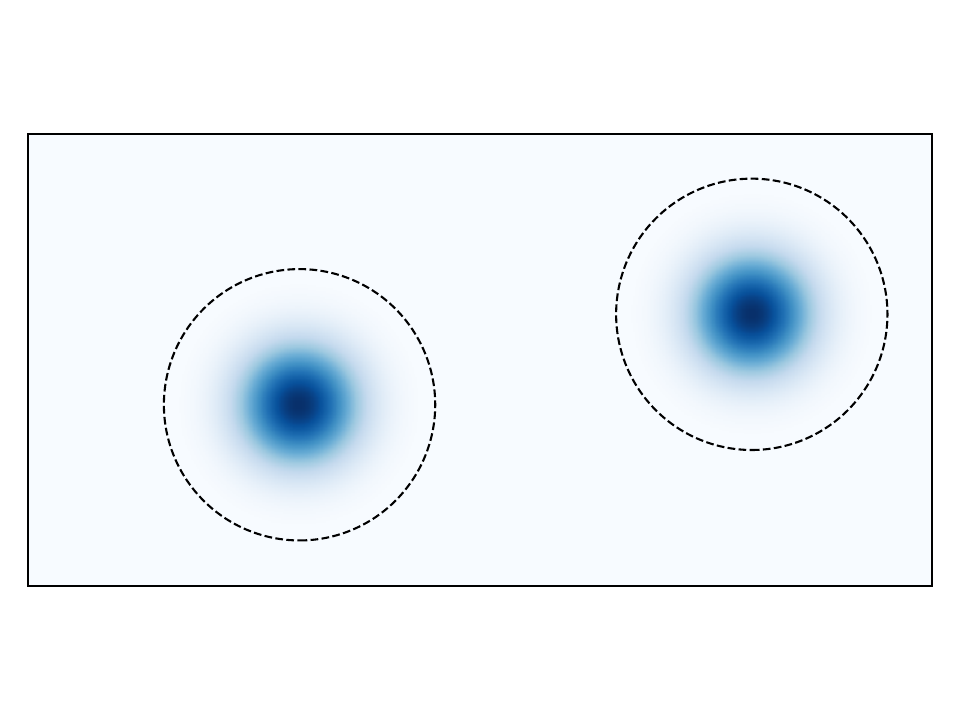}
	\includegraphics[width=\columnwidth]{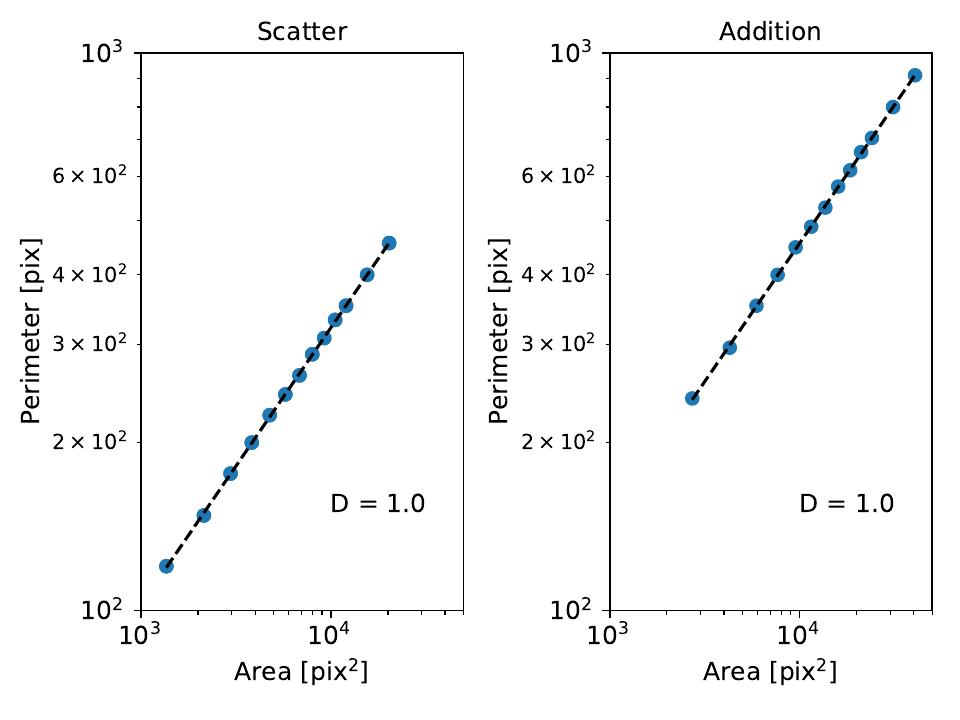}
    \caption{Two identical, non-overlapping, spherically-symmetric blobs, which yield $D_p=1$ in both interpretations of the perimeter-area method. The blobs were generated using the cubic spline kernel function shown in equation~\ref{eq:cubic-spline-3d}, and their sizes are indicated by the dashed circles.}
    \label{fig:blobs-same}
\end{figure}

\begin{figure}
	\includegraphics[width=\columnwidth]{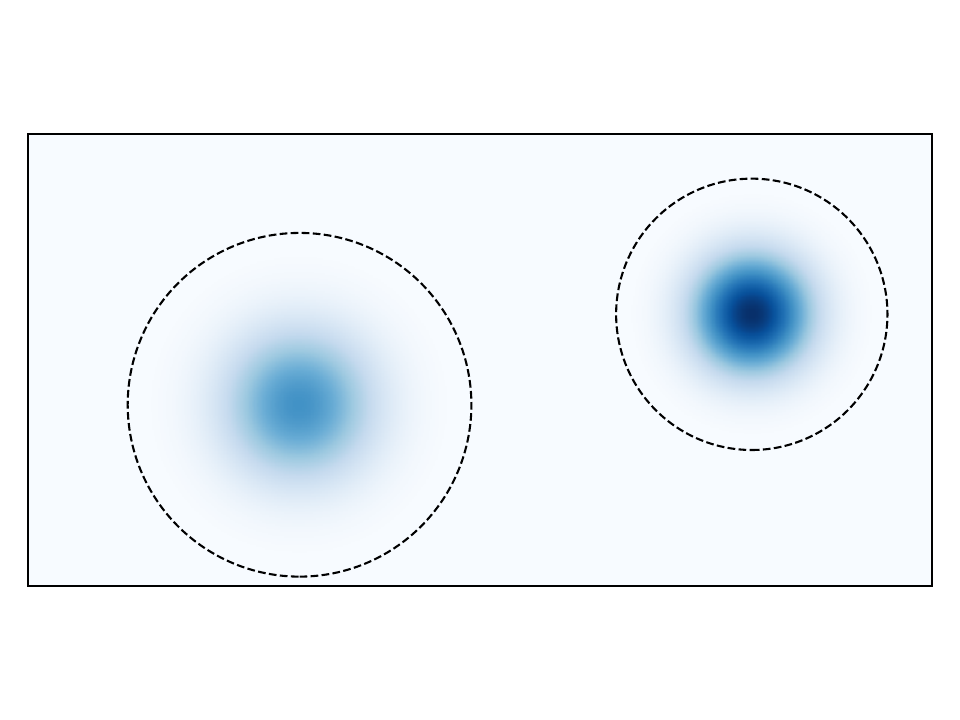}
    \includegraphics[width=\columnwidth]{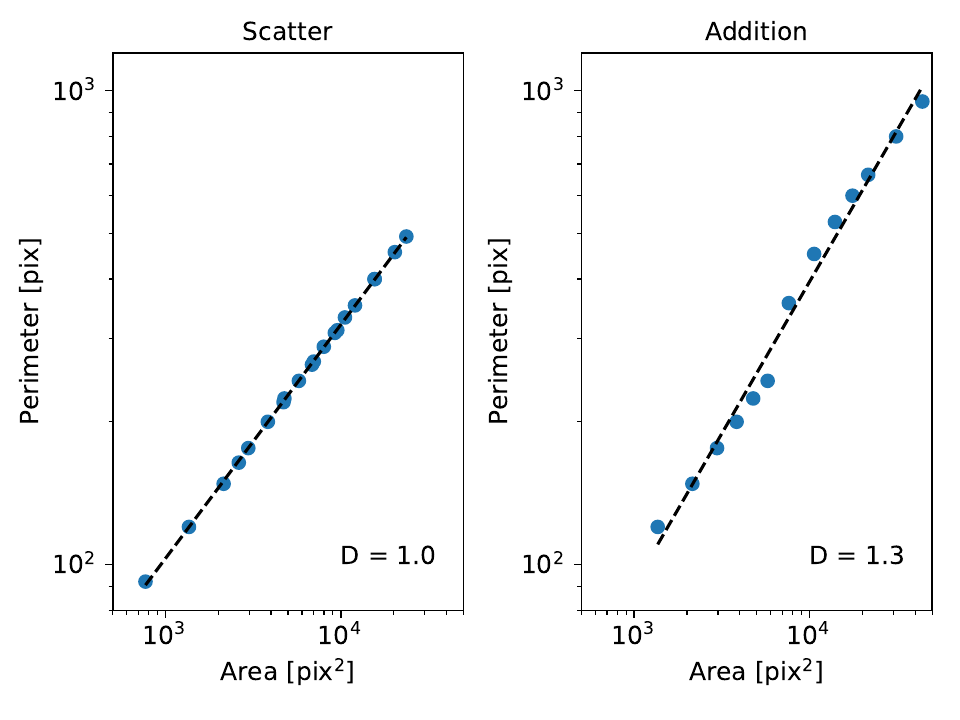}
    \caption{Two non-identical, non-overlapping, spherically-symmetric blobs, which yield $D_p=1$ in the scatter interpretation and $D_p=1.3$ in the addition interpretation. The blobs were generated using the cubic spline kernel function shown in equation~\ref{eq:cubic-spline-3d}, and their sizes are indicated by the dashed circles.}
    \label{fig:blobs-diff}
\end{figure}

Let us consider a single, finite, spherically-symmetric blob of gas. Any chosen contour of the blob is perfectly circular, and hence $P\propto A^{1/2}$. If we are to measure the fractal dimension using the perimeter-area method, we will obtain the same data set for both interpretations of the method (see Section~\ref{sec:perim-area}), since we have a single blob. This will result in a measured fractal dimension of $D_p=1$.

If we now introduce a second blob, identical to the first one and non-overlapping with it, the scatter interpretation will yield $D_p=1$. The addition interpretation will have a different set of data points $\{(2A_1,2P_1), (2A_2,2P_2), ..., (2A_n,2P_n)\}$, where $A_i$ and $P_i$ are the areas and perimeters of different contours of one of the two blobs. This data can also be fitted with a line of slope 0.5, and a fractal dimension of 1, however the constant term of the liner fit in log-log space will be different than the one in the scatter interpretation (see Figure~\ref{fig:blobs-same}). This agreement between the two methods can be generalised for a cloud consisting of N identical, non-overlapping blobs.

Finally, if we consider two spherically-symmetric but non-identical blobs, the scatter interpretation once again finds $D_p=1$. However, the data points of the addition interpretation are no longer simple multiples of the data points of the scatter interpretation, as they were in the previous case. This leads to a disagreement between the results of the two methods (see Figure~\ref{fig:blobs-diff}). Note that in the addition plot we can identify two linear segments with a slope of 0.5, but the liner fit through all of the data points has a steeper slope. 

The above two examples both serve to ask a simple question. What is the fractal dimension of a cloud which consists of two (many) independent regions, both (all) of which have the same individual fractal dimension? The scatter interpretation deduces that the fractal dimension of the cloud is always the same as the (identical) fractal dimensions of its substructures. The addition interpretation is inconclusive. In our two examples we get a fractal dimension of either 1 or 1.3, depending on the relative sizes of the cloud substructures. Due to this behaviour of the addition interpretation, we recommend the use of the scatter interpretation in future studies.

%%%%%%%%%%%%%%%%%%%%%%%%%%%%%%%%%%%%%%%%%%%%%%%%%%

% Don't change these lines
\bsp	% typesetting comment
\label{lastpage}
\end{document}